\documentclass[twocolumn]{aastex63}

\usepackage{amsmath}

\usepackage{times}

\usepackage{graphicx}

\usepackage[T1]{fontenc}
\usepackage{aecompl}
\usepackage{calligra}
\DeclareMathAlphabet{\mathcalligra}{T1}{calligra}{m}{n}
\DeclareFontShape{T1}{calligra}{m}{n}{<->s*[2.2]callig15}{}

\definecolor{colordisk}{rgb}{0.78,0.918,0.898}
\definecolor{color3}{rgb}{0.749,0.506,0.1765}

\usepackage{hyperref}
\hypersetup{
    pdfnewwindow=true,      
    colorlinks=true,       
    linkcolor=color3,      
    citecolor=color3,        
    filecolor=magenta,      
    urlcolor=color3         
}

\def\apjl{ApJL}               
\def\apjs{ApJS}               
\def\aap{A\&A}                

\defcitealias{DL:2016}{DL16}

\usepackage{color}
\usepackage{xcolor}

\newcommand{\OK}[1]{\textcolor{red}{[OK]}}

\def\eg{{\em e.g.}}
\def\ie{{\em i.e.}}
\def\eb{e_\mathrm{b}}
\def\rcav{r_\mathrm{cav}}
\def\Rcav{R_\mathrm{cav}}

\def\ab{a_\mathrm{b}}
\def\Mb{M_\mathrm{b}}
\def\ob{\Omega_\mathrm{b}}
\def\rin{r_\mathrm{in}}
\def\Rin{R_\mathrm{in}}

\def\rt{r_\mathrm{t}}
\def\rs{r_\mathrm{s}}

\def\rout{r_\mathrm{out}}
\def\css{c^2_\mathrm{s}}
\def\wo{\omega_0}
\def\wp{\omega_P}
\def\wq{\omega_Q}
\def\ss{\zeta}

\def\omecc{\omega_\mathrm{ecc}}
\def\omg{\omega_\mathrm{g}}
\def\omp{\omega_\mathrm{p}}

\usepackage{xifthen}
\definecolor{red1}{rgb}{0.7, 0.15, 0.15}
\newcommand*{\needcite}[1]{
    \ifthenelse{\equal{#1}{}}{
        {\color{red1}[?]}
    }{
        {\color{red1}[#1]}
    }
}

\usepackage{lineno}

\usepackage[makeroom]{cancel}

\begin{document}

\shorttitle{Analytical Theory of Circumbinary Disks}
\shortauthors{Grcić, D'Orazio, Pessah}
\title{Insights from Analytical Theory of Eccentric Circumbinary Disks}

\author[0000-0001-5301-2564]{Marcela Grcić}
\affiliation{Niels Bohr International Academy, Niels Bohr Institute, Blegdamsvej 17, DK-2100 Copenhagen Ø, Denmark}
\email{marcela.grcic@nbi.ku.dk}

\author[0000-0002-1271-6247]{Daniel J. D'Orazio}
\affiliation{Space Telescope Science Institute, 3700 San Martin Drive, Baltimore, MD 21218, USA}
\email{dorazio@stsci.edu}
\affiliation{Niels Bohr International Academy, Niels Bohr Institute, Blegdamsvej 17, DK-2100 Copenhagen Ø, Denmark}

\author[0000-0001-8716-3563]{Martin E. Pessah}
\affiliation{Niels Bohr International Academy, Niels Bohr Institute, Blegdamsvej 17, DK-2100 Copenhagen Ø, Denmark}
\affiliation{School of Natural Sciences, Institute for Advanced Study, 1 Einstein Drive, Princeton, NJ 08540, USA}
\email{mpessah@nbi.ku.dk}

\begin{abstract} 
Eccentric cavities in circumbinary disks precess on timescales much longer than the binary orbital period. These long-lived steady states can be understood as trapped modes in an effective potential primarily determined by the binary quadrupole and the inner-disk pressure support, with associated frequencies $\omega_Q$ and $\omega_P$. Within this framework, we show that the ratio $\omega_P/\omega_Q$ is the main parameter determining the mode spectrum, and obtain a thorough understanding of it by systematically solving this problem with various degrees of sophistication. We first find analytical solutions for truncated power-law disks and use this insight in disks with smooth central cavities. Our main findings are: (i) The number of modes increases for thinner disks and more-equal-mass binaries. (ii) For 2D disks, the normalized ground-mode frequency, $\omega_0/(\omega_Q+\omega_P)$, decreases monotonically with the ratio $\omega_P/\omega_Q$. (iii) For thin disks, $\omega_P\ll\omega_Q$, the ground-mode frequency coincides with the maximum of the effective potential, which tracks the gravitational quadrupole frequency inside the inner-disk cavity, and is thus rather sensitive to the density profile of the cavity, where these modes are localized. (iv) For thick disks, $\omega_P\gg\omega_Q$, increasing pressure support anchors the peak of the effective potential at the inner cavity radius as the ground-mode extends farther out and its frequency decreases. (v) In agreement with numerical simulations, with $\omega_P/\omega_Q \simeq 0.1$, we find that disk precession is rather insensitive to the density profile and ground-mode frequencies for 3D disks are about half the value for 2D disks.
\end{abstract}

\section{Introduction}

Gaseous circumbinary disks (CBDs) form in a range of astrophysical scenarios, and can play a large role in binary orbital evolution and electromagnetic emission from such systems. We observe CBDs in the form of gas-rich protostellar disks around young stellar binaries \citep{Beckwith_1990,Dutrey_1994,Dutrey_plus_1998,Pierens_2012,Avenhaus_Quanz_2017,Czekala_2019,Keppler_2020}, which shape stellar binary orbits \citep{Valli+2024} and the planets that surround them \citep{Hwang_El-Bardy_Rix_2022}. We also expect CBDs around massive black hole binaries (MBHBs), which form as a consequence of galaxy mergers \citep{Begelman_1980}. If the merging galaxies are gas-rich, an accretion disk may form around the MBHB \citep{Barnes:1996, Barnes:2002}, altering its path towards merger \citep[\eg,][]{Armitage_Natarajan_2002, Siwek_Kelley_Hernquist:2024} and rendering it discoverable in the electromagnetic (EM) spectrum \citep{DOrazioCharisi:2023}.

The rate-of-change of the binary orbital elements caused by the interaction with a gas disk, and the modulation of EM emission from an accreting binary are both closely connected to the response of the disk to the binary potential. In many scenarios, eccentricity is excited in the CBD. The CBD eccentricity and its precession properties are deeply intertwined with orbital evolution and observational signatures, motivating further investigation of eccentric CBD morphology.

Numerical hydrodynamical simulations of 2D and 3D CBDs show that the CBD evolves to an eccentric state over time, even if the binary's orbit is circular \citep{MacFadyen_Milosavljevic_2008,Moody_2019, KITP_CC:2024}. This leads to several sources of periodic accretion variability that depend on the eccentric disk's precession frequency, structure, and eccentricity \citep[\eg,][]{MacFadyen_Milosavljevic_2008, Shi+_2012, Dorazio_Haiman_2013, Dan-gaps, Zrake_Tiede_MacF_Haiman_2021, Noble+_2021}. For eccentric binaries, disk eccentricity can also be excited, and the disk-precession rate relative to the binary eccentricity vector mediates the ratio of mass transfer onto the binary components, allowing preferential accretion, even for equal mass binaries \citep{Dunhill+2015, MunozLai_PulsedEccAcc:2016, Siwek+_2023, binlite+2024}. Furthermore, the evolution of the binary separation and eccentricity exhibits a change when the disk transitions from a highly eccentric precessing state to a barely eccentric non-precessing state \citep{Miranda_Munoz_Lai_2017, Dan-transition, Siwek+_2023}. Similarly, \cite{Dittmann_Ryan_2022} discuss the effect of the disk aspect ratio and viscosity on torques through changes in the disk morphology and dynamics. \cite{Wang_Bai_Lai_2023} found that slow cooling in the disk can cause the disk cavity to become circular, followed by suppressed accretion variability of the binary, and an increased specific angular momentum transferred from the disk to the binary.

To probe the development of this important aspect of circumbinary disk morphology, and to gain further insight into binary-disk dynamics, we employ semi-analytical techniques to investigate the disk eccentricity and precession frequency. Such methods utilize perturbation theory to describe the deviation of thin disks from an axisymmetric state \citep{Lubow_1991,Ogilvie_2001, GO2006}. Such perturbative models provide interpretation of numerical solutions to the full equations of hydrodynamics, and extend beyond the numerically accessible range of parameter space.

Previously, a number of works have applied this approach to interpret results from numerical hydrodynamical calculations of the full non-linear fluid flow for CBDs. \cite{Miranda_Munoz_Lai_2017} used this approach to identify possible causes for a changing disk eccentricity and precession frequency with binary mass ratio and eccentricity. \cite{Shi+_2012} and \cite{Siwek+_2023} employed such semi-analytical methods to find a good match to the computed disk eccentricity distribution and precession frequency. \cite{ML2020} carried out a study of locally isothermal CBDs and found that the precession rates derived from the perturbative approach match reasonably well with numerical hydrodynamical simulations, at least in the tested regions of parameter space. 
 
Applying the perturbative method for the disk eccentricity to CBDs generally results in a consistent picture. Namely, an eccentric, freely precessing disk is a natural result of a centrally truncated disk density profile \citep{ML2020}; the disk is significantly eccentric in its innermost part \citep{Ogilvie_2001,GO2006,LeeDempLith_LongLived:2019,ML2020}; the binary gravitational potential drives, and the pressure in the disk damps the disk precession frequency; multiple solutions for the disk precession rate and eccentricity are possible in thinner disks \citep{LeeDempLith_LongLived:2019,ML2020}; and non-precessing, eccentric disk solutions exist and were observed in numerical simulations \citep{2008Planetesimal, Siwek+_2023, binlite+2024}. Additionally, \cite{Lubow2022} proposes a resonant-like increase in the eccentricity of non-precessing disks around eccentric binaries for suitable values of the disk aspect ratio, but this phenomenon has not yet been seen in numerical hydrodynamical calculations.

While perturbative methods have shed light on several aspects of the CBD problem, its applications are somewhat limited. The limitations arise because we require knowledge of the background non-eccentric disk solutions to predict the perturbed eccentric state. Different studies adopt different choices to model the CBD, making systematic comparisons difficult. This in turn complicates the predictive use of the perturbative approach.

Choices for the binary model are mostly straightforward, and include the expansion of the binary potential and its parameterization in binary mass ratio, separation, and eccentricity. The disk model, however, can be less straightforward and includes the disk dimension, the gas equation of state, and importantly, a choice for the radial density distribution. Therefore, it is crucial to know how much, and in what way, the assumptions underlying the disk model influence disk eccentricity solutions.

The aim of this paper is to provide a coherent framework to exploit the insights obtained from a perturbative, semi-analytical approach in a systematic way. To achieve this, we conduct an analysis over a range of disk density models and system parameters. We find steady-state solutions for the disk eccentricity and precession rate given increasingly complex density profiles to explain and quantify how some aspects of steady-state eccentricity solutions depend on the density profile. This is facilitated by starting with disk models that allow analytical analyses, thus providing detailed insights into results.

We divide the remaining part of this paper into four sections. In Section \ref{sec:methods}, we provide a brief overview of the equations that describe the evolution of the eccentricity in a circumbinary disk, and present the methods we use to analyze and solve those equations. In Section \ref{sec:3}, we assume a truncated power-law-density radial profile. First, in \S\ref{subsec: Disk Density Profiles Allowing Trapped Modes}, we discuss density profiles resulting in an eccentricity potential that allows for trapped modes. Then, in \S \ref{sec:3a}, we find analytic solutions for the eccentricity of non-precessing disks. Then, in \S \ref{sec:3b}, we further the analytic treatment to find the number of possible disk eccentricity modes. In Section \ref{sec: Disk Density Impact on Eccentricity}, we adopt radial density profiles including smooth central cavities, and a mass and angular momentum transfer from the disk to the binary. For these density profiles, we find numerical solutions for the perturbative eccentricity equations to identify the solution properties that are insensitive to the choice of the density model. We end with a summary of the relevant results and a discussion in Section \ref{sec: Discussion and conclusion}.

\section{Methods}\label{sec:methods}

\begin{figure}[h]
    \centering
\includegraphics{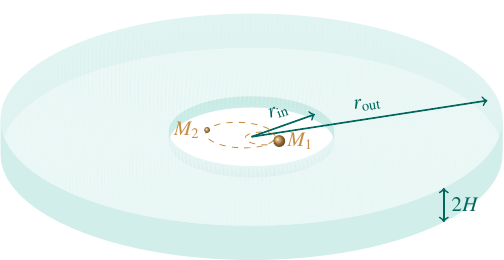}
    \caption{The binary-disk system. We illustrate two masses, $M_1$ and $M_2$ (solid brown), their orbits (dashed brown lines), and a co-planar circumbinary disk (light green). The disk size is determined by its inner radius $\rin$, outer radius $\rout$, and the disk scale height $H(r)$.}
    \label{fig: an illustration of the system}
\end{figure}

We consider a binary surrounded by a thin, co-planar circumbinary disk. The binary's eccentricity, semi-major axis, mass, mass ratio, and orbital frequency are $\eb$, $\ab$, $\Mb\equiv M_1+M_2$, $q\equiv M_2/M_1\leq 1$, $\ob=\sqrt{G\Mb/\ab^3}$, where $M_1$ and $M_2$ are the masses of the primary and the secondary. The gaseous disk has an inner radius $\rin$, outer radius $\rout$, and its scale height is $H(r)$. In Figure \ref{fig: an illustration of the system}, we illustrate this system, and its characterizing quantities.

\subsection{The Eccentricity Equation}\label{subsection:The Eccentricity Equation}

To describe the eccentricity evolution of a circumbinary disk, we follow \cite{GO2006} and \cite{TO2016}, who derived perturbed equations of hydrodynamics for fluid elements moving under the influence of the pressure in the disk and the gravitational potential of the binary. In this framework, we write the values of the quantities of the disk fluid elements (density, pressure, velocity, and temperature) as sums of unperturbed and perturbed values $X_\mathrm{tot}=X+X_\mathrm{per}$ \citep{Lee_Dempsey_Lithwick_2019}. The reference (unperturbed) state of the disk is time-independent and azimuthally symmetric, described by surface density and pressure profiles $\Sigma$ and $p$. Additionally, the fluid elements in the reference state are on fixed circular orbits (with some angular velocity $v$ and no radial velocity $u=0$). As a measure of perturbation from that reference state, we define the complex eccentricity $E=e\exp{(i\overline{\omega})}$ \citep{GO2006}, with eccentricity magnitude $e$ and the argument of the longitude of pericenter of the fluid streamlines $\overline{\omega}$ \citep{Armitage_2008}. Specifically, $E(r,t)$ is proportional to the radial ($\Tilde{u}$)  and angular ($\Tilde{v}$) velocity perturbation: $ \Tilde{u} \approx ir\Omega E(r)$, and $\Tilde{v} \approx r\Omega E(r)/2$, where $\Omega=\sqrt{G\Mb/r^3}$ is the orbital frequency of a disk fluid element at distance $r$ from the binary center of mass. The perturbed quantities are connected to $m=1$ Fourier coefficients $\Tilde{X}$ as $X_\mathrm{pert}(r,\phi,t)=\Re \left[\Tilde{X}(r)e^{i(\phi-\omega t)} \right]$ \citep{Lee_Dempsey_Lithwick_2019}.

We then obtain a differential equation for the evolution of the complex eccentricity function $E(r,t)$ \citep{GO2006}:
\begin{equation}\label{eq: eccentricity equation in terms of f_p and f_g}
2r^3\Sigma\Omega\partial_t E(r,t)=if_\mathrm{p}(E,r)+if_\mathrm{g}(E,r),
\end{equation}
where $f_\mathrm{g}$ and $f_\mathrm{p}$ are eccentricity forces that determine the disk eccentricity distribution evolution due to gravitational and pressure effects, respectively. We denote radial and time derivatives as $\partial_r\equiv \partial/\partial r$ and $\partial_t\equiv \partial/\partial t$.

Throughout, we consider systems that have reached a quasi-steady-state, in which the (time-independent) radial eccentricity distribution $E(r)$ precesses with a constant frequency $\omega$ and thus:
\begin{equation}
\label{eq: steady state time derivative of E(r,t)}
\partial_tE(r,t)=i\omega E(r,t)=i\omega E(r)e^{i\omega t}.
\end{equation}
 In general, for a given set of boundary conditions, the temporal dependence we assumed in Eq. ~(\ref{eq: steady state time derivative of E(r,t)}) leads to an eigenvalue problem for the frequency $\omega$, where $E(r)$ is the associated (complex) time-independent eccentricity eigenmode associated with $\omega$. We discuss this further in \S \ref{subsection: Schrödinger form of the eccentricity equation and the eccentricity potential} and \S \ref{S:Methods_numerics}.
 
Additionally, in Eq. (\ref{eq: eccentricity equation in terms of f_p and f_g}), for simplicity, we neglect the effects of the eccentricity forcing (such as Lindblad resonances) and damping (such as viscosity). For example, \cite{ML2020} argue that whatever the details of such effects are, they cause the time evolution of the eccentricity amplitude $E(r)$, so they must balance each other out by the time the disk achieves a steady state (Eq. (\ref{eq: steady state time derivative of E(r,t)})). 

The gravitational term due to the time-averaged quadrupole gravitational potential $\Phi_2$ of the binary is:
\begin{equation}\label{eq: gravitational forcing term for the eccentricty equation}
    f_\mathrm{g}\equiv -r\Sigma \partial_r\left[r^2 \partial_r\Phi_2 \right]E=\Omega^2 \Sigma \ab^2r QE.
\end{equation}
Here, we have included the lowest order binary eccentricity term in the quadrupole gravitational potential $\Phi_2$ \citep{Goldreich_Tremaine_1980,H2009,Miranda_Lai_2015} to write the last equality in Eq. (\ref{eq: gravitational forcing term for the eccentricty equation}), and to define $Q$ as a measure of the effect of the binary gravitational quadrupole potential:
\begin{equation}\label{eq: definition of the Q parameter}
    Q\equiv\frac{3q}{2(1+q)^2}\left(1+\frac{3}{2}\eb^2 \right) .
\end{equation}
If the binary is eccentric, additional forcing should be included \citep{ML2020} so in this work, we assume a  circular binary ($\eb=0$).

\subsection{Thermodynamic Effects in Pressure Forcing}

The pressure forcing effect is more complex, as it depends on whether we choose a 2D or a 3D treatment of the disk, and on the character of the allowed perturbations to the steady circular motion of the disk element. In the 2D treatment, there is no vertical motion of the fluid element in the disk, whereas in the 3D treatment, small perturbations to the initially zero vertical velocity are allowed. \cite{Ogilvie_2001} developed a formalism for the eccentricity equation for adiabatic disks, whereas \cite{TO2016} developed it for locally isothermal disks. As pointed out in \citet{TO2016}, the adiabatic eccentricity equation reduces to their locally isothermal version (where the adiabatic index is $\gamma=1$) only when the sound speed $\css$ is a global constant. This is a rather stringent limitation and it is not met in the case of, \eg, commonly modeled disks with a constant disk aspect ratio where $\css = \gamma P/\Sigma \propto 1/r$. 
 
To monitor the effects that the thermodynamics of the perturbations have on the eccentricity solutions, we introduce a label-parameter $\xi$, which can take the value of either 1 or 0. This allows us to keep track of how the character of the temperature perturbations propagates through the eccentricity equation and its solutions throughout the rest of this work. Setting $\xi =1$ allows for the temperature perturbation required for fluid elements to be on eccentric isentropic orbits. Setting $\xi=0$ eliminates all temperature perturbation, and results in a fixed radial temperature profile (locally isothermal disk). We use the aforementioned definition of $\xi$, and results obtained by \cite{TO2016}, to write the pressure term as: 
\begin{subequations}\label{eq: pressure forcing term f_p for the eccentricty equation}
\begin{align}
    \label{eq22}
    \begin{split}
f^\mathrm{2D}_\mathrm{p}=&Er^2  \partial_rp+\partial_r\left[ p  r^3\partial_rE \right]\gamma^\xi\\ &-\partial_r\left[\Sigma \partial_r\css  r^3E \right]\left(1-\xi\right),
\end{split}\\
    \label{eq44}
\begin{split}
f^\mathrm{3D}_\mathrm{p}=&Er^2\partial_r p \left(4-\frac{3}{\gamma} \right)^\xi+\partial_r\left[pr^3\partial_rE \right]\left(2-\frac{1}{\gamma}\right)^\xi\\
& +6prE\left(\frac{\gamma+1}{2\gamma} \right)^\xi \\
&-\partial_r\left[\Sigma \partial_r\css E \right]r^3\left(1-\xi\right) ,
    \end{split}
\end{align}
\end{subequations}
where the superscripts indicate the disk dimension, and $\Sigma$ and $p$ are vertically averaged surface density and pressure. 
The choice of $\xi=0$, $\gamma=1$ corresponds to the locally isothermal equations in \cite{TO2016}. The latter case is also the equation used in \cite{ML2020} and the form we use throughout the rest of this paper. Setting $\xi=1$, and $\gamma>1$ provides the equations for studying eccentric disks subject to isentropic perturbations.

\subsection{Eccentricity Boundary Value Problem}\label{subsection: Eccentricity boundary value problem}

Equations (\ref{eq: eccentricity equation in terms of f_p and f_g})-(\ref{eq: pressure forcing term f_p for the eccentricty equation}) amount to a linear homogeneous second order ordinary differential equation for $E(r)$ (Appendix \ref{app B - general eccentricty potential and bessel calculations}, Eq. (\ref{z1})):
\begin{equation}\label{eq: eccentricty equation in its standard E''+xE'+yE=0 form}
        E''(r)+f_{E'}(r)E'(r)+f_{E}(r)E(r)=0,
\end{equation}
whose  general solution is a linear combination of two linearly independent solutions $E_1(r)$ and $E_2(r)$:
\begin{equation}\label{eq: eccentricity profile solution in its E=b_1E_1+b_2E_2 form}
E(r)=\beta_1E_1(r)+\beta_2E_2(r),
\end{equation}
where $\beta_1$ and $\beta_2$ are constants whose values depend on the boundary conditions. The expressions for the functions $f_{E'}$ and $f_{E}$ are provided in Appendix \ref{app B - general eccentricty potential and bessel calculations}.

Following \citet{Ogilvie_2001} and \citet{2008Planetesimal}, we demand the Lagrangian pressure perturbations to vanish at the disk's inner, $\rin$, and outer, $\rout$, radii:
\begin{equation}\label{eq: the boundary condition - general}
    \left. \left[ \frac{E}{(p/\Sigma)^{1-\xi}} \right]' \right|_{\rin}=\left. \left[ \frac{E}{(p/\Sigma)^{1-\xi}} \right]' \right|_{\rout}=0.
\end{equation}

Note that setting $\xi=1$ into Eq. (\ref{eq: the boundary condition - general}) recovers the adiabatic eccentricity boundary condition $\partial_r E(\rin)=\partial_r E(\rout)=0$ \citep{Ogilvie_2001}, whereas setting $\xi=0$, Eq. (\ref{eq: the boundary condition - general}) leads to eccentricity boundary condition for locally isothermal disks $\partial_r \left(E/\css \right)(\rin)=\partial_r \left(E/\css \right)(\rout)=0$, where $c_\mathrm{s}$ is the isothermal sound speed.

Substituting $E(r)$ from Eq. (\ref{eq: eccentricity profile solution in its E=b_1E_1+b_2E_2 form}) into Eq. (\ref{eq: the boundary condition - general}) allows us to find the relationship between the constants $\beta_1$ and $\beta_2$:
\begin{subequations}\label{eq: solutions for constants beta_1 and beta_2 in terms of E_1 and E_2}
\begin{align}
    \label{eq: trivial constants beta_1 and beta_2}
        \beta_1&=\beta_2=0  &;
         \qquad \Psi_{\rm BC}\neq0, \\
    \label{eq: non-trivial constants beta_1 and beta_2  in terms of E_1 and E_2}
        \frac{\beta_2}{\beta_1}&=-\left.\frac{\left[E_{1}\left(\frac{\Sigma}{p}\right)^{1-\xi}\right]'}{\left[E_{2}\left(\frac{\Sigma}{p}\right)^{1-\xi}\right]'}\right|_{\rin,\rout} &;
         \qquad \Psi_{\rm BC}=0.
\end{align}
\end{subequations}
where the relationship between the pressure perturbations at the two boundaries (the determinant of the system of equations for $\beta_1$ and $\beta_2$ given by Eq. (\ref{eq: the boundary condition - general})):
\begin{equation}
\label{eq: zeta definition}
\begin{split}
    \Psi_{\rm BC}\equiv& 
    \left.\left[E_1\left(\frac{\Sigma}{p}\right)^{1-\xi}\right]'\right|_{\rin}
    \left.\left[E_2\left(\frac{\Sigma}{p}\right)^{1-\xi}\right]'\right|_{\rout}\\ 
  &-\left.\left[E_{2}\left(\frac{\Sigma}{p}\right)^{1-\xi}\right]'\right|_{\rin}
    \left.\left[E_{1}\left(\frac{\Sigma}{p}\right)^{1-\xi}\right]'\right|_{\rout},
    \end{split}
\end{equation}
determines the existence of non-trivial solutions. By Eq. (\ref{eq: trivial constants beta_1 and beta_2}), if $\Psi_{\rm BC} \neq 0$, the eccentricity solution (Eq. (\ref{eq: eccentricity profile solution in its E=b_1E_1+b_2E_2 form})) is trivial; $E(r)=0$, and a non-trivial eccentricity solution (Eq. (\ref{eq: eccentricity profile solution in its E=b_1E_1+b_2E_2 form})) exists only if the condition $\Psi_{\rm BC}=0$ is satisfied.

\subsection{Schrödinger's Eccentricity Equation and the Eccentricity Potential}
\label{subsection: Schrödinger form of the eccentricity equation and the eccentricity potential}

In order to provide more insight into the eccentricity equation (Eqs. (\ref{eq: eccentricity equation in terms of f_p and f_g})-(\ref{eq: pressure forcing term f_p for the eccentricty equation})) and its solutions, we write it in the form of a Schrödinger equation \citep{Ogilvie:2008,Lee_Dempsey_Lithwick_2019, LeeDempLith_LongLived:2019,ML2020}:
\begin{equation}
\label{eq: Schrödinger equation for y}
    y''(r)+k^2(r)y(r)=0,
\end{equation}
for the scaled eccentricity $y(r)$ (Eq. (\ref{eq: scaled ecc})):
\begin{equation}\label{eq: scaled eccentricty y for Schrödinger equation}
y=Ep^{\xi/2}r^{3/2}\Sigma^{\left(1-\xi\right)/2}.
\end{equation}
The radial wavenumber $k(r)$ satisfies:
\begin{equation}
\label{eq: wave number k^2 for Schrödinger equation}
k^2 \equiv \frac{2 r^{-2}}{h_e^2 \ob}\left(\frac{r}{\ab}\right)^{3/2}\left[\omecc -\omega \right],
\end{equation}
where the effective disk scale height $h_e$ is either 
$h^{\mathrm{2D}}_e=h\sqrt{\gamma}$ or
$h^{\mathrm{3D}}_e=h\sqrt{(2\gamma-1)/\gamma}$, where $h$ is the disk aspect ratio for a locally isothermal disk:
\begin{equation} \label{eq: disk aspect ratio definition}
    h\equiv \frac{H}{r}=\sqrt{\frac{ p}{\Sigma r^2 \Omega^2}}.
\end{equation}
Above, we introduced the eccentricity potential $\omecc$:
\begin{equation}
\label{eq: eccentricity potential as a sum of its pressure and gravitational parts}
    \omecc \equiv \omg + \omp,
\end{equation}
as the sum of the gravitational eccentricity potential $\omega_\mathrm{g}$:
\begin{equation}\label{eq: gravitational part of the eccentricity potential}
  \omg\equiv \frac{Q}{2}\left(\frac{r}{\ab}\right)^{-7/2}\ob,
\end{equation}
and the pressure eccentricity potential $\omega_\mathrm{p}$:
\begin{subequations}\label{eq: pressure part of the eccentricity potential}
    \begin{align} \label{eq: 2D pressure part of the eccentricity potential}
    \begin{split}
        \omega^{\mathrm{2D}}_\mathrm{p}\equiv & \frac{h^2_e}{2}\ob \left(\frac{r}{\ab}\right)^{-3/2}\left[\frac{rp'}{p\gamma^\xi} -\frac{r^2\left(r^3\Sigma^{1-\xi}p^\xi\right)''}{2 r^3\Sigma^{1-\xi}p^\xi}       \right.\\
        & +\left( \frac{r\left(r^3\Sigma^{1-\xi}p^\xi\right)'}{2 r^3\Sigma^{1-\xi}p^\xi}\right)^2\left. -\frac{1-\xi}{rp}\left[\Sigma c_\mathrm{s}^{2'}r^3\right]'\right],
        \end{split}\\
         \label{eq: 3D pressure part of the eccentricity potential}
        \begin{split}
        \omega^{\mathrm{3D}}_\mathrm{p}\equiv &\frac{h^2_e}{2}\ob \left(\frac{r}{\ab}\right)^{-3/2}\left[ -\frac{r^2\left(r^3\Sigma^{1-\xi}p^\xi\right)''}{2 r^3\Sigma^{1-\xi}p^\xi}  \right.\\
            &+\left( \frac{r\left(r^3\Sigma^{1-\xi}p^\xi\right)'}{2 r^3\Sigma^{1-\xi}p^\xi}\right)^2 + \frac{rp'}{p}\left( \frac{4\gamma-3}{2\gamma-1}\right)^{\xi}\\
            &\left. +6\left(\frac{\gamma+1}{2(2\gamma-1)} \right)^\xi-\frac{(1-\xi)r^2}{p}\left[\Sigma (\css)'\right]'\right] .
        \end{split}
    \end{align}
\end{subequations}
The advantage of writing the eccentricity equation in a Schrödinger's form lies in exploitable quantum mechanics results. In analogy with the quantum mechanics wave equation, from Eq. (\ref{eq: Schrödinger equation for y}), we see that the scaled eccentricity $y(r)$, and thus $E(r)$, can oscillate only at radii where $k^2>0$. By Eq. (\ref{eq: wave number k^2 for Schrödinger equation}), $k^2>0$ is only satisfied where the precession frequency of the disk $\omega$ is smaller than the eccentricity potential $\omecc$.

\subsection{Solution Methods}
\label{S:Methods_numerics}

Generally, the factor determining the existence of solutions, $\Psi_{\rm BC}$ (Eq. (\ref{eq: zeta definition})), is a function of the gas properties ($p$, $\Sigma$, $\gamma$), the disk size ($\rin$, $\rout$, $h$), the binary parameter $Q$, and the precession frequency $\omega$. 
We solve the eccentricity boundary value problem to find the value of the parameter of interest, most often $\omega$, and the associated eccentricity profile $E(r)$. Depending on the system set-up (\eg, density profile), different methods of solving the boundary value problem prove most useful.

{\it Analytic } \/-- In some cases (\S \ref{sec:3a}), an analytic solution (Eq. (\ref{eq: eccentricity profile solution in its E=b_1E_1+b_2E_2 form})) for the eccentricity equation exists. If it does, we can find analytic expressions for the $\Psi_{\rm BC}=0$ condition (Eq. (\ref{eq: zeta definition})). Analytic solutions for the eccentricity profile and eigenvalues in a simple form are possible only for a small subset of possible binary-disk systems. We look into some of those cases in Section \ref{sec:3}. For other systems, for which analytic solutions are not possible, we turn to numerical methods.

{\it Numerical methods} \/-- In \S\ref{sec: Disk Density Impact on Eccentricity}, we supplement analytical analyses with numerical solutions to the boundary value problem.
We implement the shooting method to solve the eigenvalue problem for $\omega$ for given values of $h$ and $Q$. The shooting method consists of solving the initial value problem for a variety of values of $\omega$ and looking for the value of $\omega$ that results in a solution that satisfies the boundary condition at the outer edge. Specifically, we use the \texttt{SciPy} \texttt{ivp} python solver to integrate the eccentricity equation (Eq. (\ref{eq: eccentricity equation in terms of f_p and f_g}) - Eq. (\ref{eq: pressure forcing term f_p for the eccentricty equation})) from its inner radius, with initial values of $E$ and $E'$ set by the boundary condition (Eq. (\ref{eq: the boundary condition - general})). We then use the \texttt{SciPy} \texttt{fsolve} function to find the value of $\omega$ required for the values of $E$ and $E'$ to satisfy the outer boundary condition once the integration is performed. The shooting method that uses the \texttt{SciPy} \texttt{fsolve} function requires an initial guess for the solution frequency for which it provides a single solution, making it impractical for multiple solution systems. To find all solutions for multiple solution systems, we still utilize the shooting method. But, instead of using the \texttt{SciPy} \texttt{fsolve} function, we find the allowed range of $\omega$ values, choose several hundreds values of $\omega$ within that range, and solve the eccentricity equation as the initial value problem for those pre-chosen $\omega$ values.  We then identify the solutions as $\omega$ values that satisfy outer boundary condition. 

{\it Schrödinger's form} \/-- We use the Schrödinger's form of the eccentricity equation (\S  \ref{subsection: Schrödinger form of the eccentricity equation and the eccentricity potential}), specifically the eccentricity potential, to complement both the analytic and numerical methods. In this way, the eccentricity potential given by Eq. (\ref{eq: eccentricity potential as a sum of its pressure and gravitational parts}) allows us to to estimate the radial extent of the disk exhibiting significant eccentricity. It also allows us to constrain the possible values of $\omega$, crucial for the implementation of numerical methods (providing a reasonable guess for \texttt{fsolve} and the frequency span when multiple modes are present). 

\section{Power-law density disks}\label{sec:3}
As a basis for the development of a better understanding of more elaborate disk models, we focus first on a disk with a constant disk aspect ratio $h$, and a power-law density profile:
\begin{equation}   
\label{eq: power-law density profile}
    \Sigma(r) =\Sigma_0 (r/\ab)^{\ss},
\end{equation}
where $\Sigma_0$ and $\ss$ are constants.

We define measures of pressure and gravitational-induced precession frequencies as:
\begin{equation}\label{eq:omega_PQ}
   \wp \equiv  \frac{h_e^2}{2} \Omega_* \qquad \textrm{and} \qquad \wq \equiv \frac{Q}{2} \left(\frac{\ab}{r_*}\right)^{2} \Omega_*,
\end{equation} 
where $\Omega_*=\sqrt{G\Mb}r_*^{-3/2}$ is the Keplerian frequency at some reference radius $r_*$. For the power-law density profile, the reasonable choice for the reference radius is the disk inner radius $\rin$, defining the edge of a sharp cavity.

We use the definitions given by Eq. (\ref{eq:omega_PQ}) to write $k^2(r)$ (Eq. (\ref{eq: wave number k^2 for Schrödinger equation})) for power-law density disks (Eq. (\ref{eq: power-law density profile})) as:
\begin{equation}\label{eq:k for power law disk}
    k^2(r)r^2=\frac{\wq}{\wp}\left(\frac{\rin}{r}\right)^2-\left(\nu^2-\frac{1}{4} \right)-\frac{\omega}{\wp}\left( \frac{r}{\rin}\right)^{3/2}.
\end{equation}
The value of $\nu$ is determined by the density power-law-exponent $\ss$, and the adiabatic index $\gamma$:
\begin{subequations}
\label{eq: nu Bessel parameter}
\begin{flalign}
\begin{split}
\left(\nu^2\right)^\mathrm{(2D)}&=\frac{\ss^2}{4}-\ss\left(\frac{2-\gamma}{2\gamma} \right)^\xi+\left(\frac{4+\gamma}{4\gamma} \right)^\xi,
\end{split}\\
\begin{split}
\left(\nu^2\right)^\mathrm{(3D)}&=\frac{\ss^2}{4}-\ss\left( \frac{6\gamma-5}{4\gamma-2}\right)^\xi-2\left( \frac{25-6\gamma}{16\gamma-8}\right)^\xi,
\end{split}
\end{flalign}
\end{subequations} 
where the superscripts 2D and 3D indicate the disk dimension.

\begin{table}[t!]
\begin{center}
\begin{tabular}{ c  | c | c | c } 
 & razor-thin  & intermediate & very thick  \\
  & disks  & (simulations) &  disks \\
 \hline
 $\wp/\wq$ & $\approx  10^{-5}$ & $\approx 0.17$ & $\approx 167$ \\
 $\left\{h, r_*/\ab, q \right\}$&  $\left\{10^{-3}, 2, 1 \right\}$ & $\left\{0.1, 2.5, 1 \right\}$ & $\left\{0.25, 2, 10^{-3} \right\}$ \\
\end{tabular}
\end{center}
\caption{Reference values of the pressure to gravity frequency ratio, where we have assumed a circular binary and $\gamma=1$. 
}
\label{table:cav_ratio}
\end{table}

From the Schrödinger form of the eccentricity equation (Eq. (\ref{eq: Schrödinger equation for y}), Eq. (\ref{eq: scaled eccentricty y for Schrödinger equation}), and Eq. (\ref{eq:k for power law disk})), we see that the precession frequency and system parameters $(Q,h)$ appear only through the ratios $\wq/\wp$ and $\omega/\wq$ (or alternatively $\omega/\wp$). Consequently, for a given density profile, as determined by $\ss$ and $\rin$, and disk type, as determined by the disk dimension and the equation of state ($\gamma$, $\xi$), the solutions for the mode frequency $\omega/\wq$ and the eccentricity profile are uniquely determined by the value of the ratio $\wq/\wp$. The same is true for different density profiles, as found by \citep{ML2020}. For this reason, throughout this paper, we frequently present different eccentricity solutions as functions of $\wq/\wp$.\footnote{We note that our definition of the pressure-induced precession frequency at the disk inner edge $\wp$ (and therefore the ratio $\wp/\wq$) differs by a factor of $2$ from the definition used by \cite{ML2020}} 

Since the ratio $\wp/\wq$ is the key dimensionless quantity determining the problem, we consider three reference values representative of razor thin disks, very thick disks, as well as intermediate disk that is representative of hydrodynamical simulations of CBDs \citep[\eg,][]{KITP_CC:2024}. These values are detailed in Table \ref{table:cav_ratio}.

\subsection{Disk Density Profiles Allowing Trapped Modes}
\label{subsec: Disk Density Profiles Allowing Trapped Modes}

As a starting point, we are interested in understanding the conditions for trapping eccentric modes in power-law disks. 
To this end, we write the gravity-induced  eccentricity potential (Eq. (\ref{eq: gravitational part of the eccentricity potential})), the pressure eccentricity potential (Eq. (\ref{eq: pressure part of the eccentricity potential})),
and the total eccentricity potential (Eq. (\ref{eq: eccentricity potential as a sum of its pressure and gravitational parts})) for power-law density disks (Eq. (\ref{eq: power-law density profile})) in terms of the gravitational and pressure-induced precession frequencies at the inner radius (Eq. (\ref{eq:omega_PQ})):
\begin{equation}\label{eq: pressure ecc potential for power law disks}
    \omp=-\wp\left(\nu^2-\frac{1}{4} \right) \left(\frac{r}{\rin}\right)^{-3/2},
\end{equation} 
\begin{eqnarray}
    \omg=\wq \left( \frac{r}{\rin}\right)^{-7/2},
\end{eqnarray}
\begin{eqnarray}\label{eq: ecc pot power}
    \omecc= \wq\left[ \left( \frac{\rin}{r}\right)^{2}-\frac{\wp}{\wq}\left( \nu^2-\frac{1}{4}\right) \right]\left( \frac{\rin}{r}\right)^{3/2}.
\end{eqnarray}

By Eq. (\ref{eq: Schrödinger equation for y}), radial oscillatory solutions are possible only in regions where $k^2(r)>0$. With that in mind, we use the eccentricity potential to constrain the range of values that mode precession frequencies $\omega$ can take. Physically viable profiles must involve a finite radial extent for the eccentricity. The solution for the frequency has to be greater than the minimum of the eccentricity potential (Eq. (\ref{eq: ecc pot power})), otherwise the eccentricity solution is a free wave\footnote{For example, see (right panel of) Figure 5 in \cite{ML2020}.}. By Eq. (\ref{eq: ecc pot power}), far away from the binary, the eccentricity potential is $\lim_{r\to \infty} \omecc= \pm 0$, where the sign depends on the density power-law-exponent (see \S \ref{subsec: Schrödinger's Insight on Eccentricity - POWER LAW}). This puts the lower constraint on the frequency as $\omega_n \gtrapprox 0$, because we exclude solutions that oscillate at infinite radius.
The upper constraint is given by requiring that $k^2(r)>0$ within the potential well, so the value of $\omega$ has to be lower than the maximum of the eccentricity potential. In other words, for a trapped mode, we require the frequency to be such to allow for the existence of turning points ($k^2=0$).  By Eq. (\ref{eq: ecc pot power}), the eccentricity potential, if positive, has a maximum at $r=\rin$.  The possible range of the mode frequencies for power-law density profiles is then: 
\begin{equation}
\label{eq: the range of trapped mode frequencies in terms of Bessel parameters - v2}
    0\lessapprox \omega \leq \wq \left[ 1-\frac{\wp}{\wq} \left( \nu^2-1/4\right)\right].
\end{equation}
In order for Eq. (\ref{eq: the range of trapped mode frequencies in terms of Bessel parameters - v2}) to be satisfied, it is necessary that $0<\wq\left[ 1-\wp/\wq \left( \nu^2-1/4\right)\right]$. In other words, knowing that $\wq>0$, we find that we can expect eccentricity modes only for:
\begin{eqnarray}\label{eq:condition on disk parameters for modes}
    \frac{\wp}{\wq}\left(\nu^2-1/4 \right)<1.
\end{eqnarray}

Equation (\ref{eq:condition on disk parameters for modes}) is always satisfied if $\nu^2-1/4<0$, which is equivalent to a positive pressure eccentricity potential (Eq. (\ref{eq: pressure ecc potential for power law disks})). For a 2D locally isothermal disk, $\nu^2-1/4<0$ is satisfied for $1< \ss<3$. For a 3D locally isothermal disk, the inequality is satisfied for $2-\sqrt{13}<\ss<2+\sqrt{13}$.

\begin{figure}[h]
    \centering
    \includegraphics[scale=1]{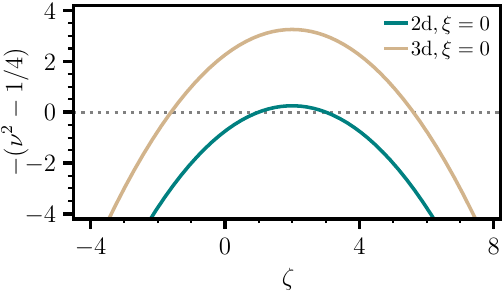}
    \caption{The magnitude of the pressure eccentricity potential $\omp/(\wp (r/\rin)^{-3/2})$ for 2D (solid dark green line) and 3D (solid brown line) locally isothermal disks for density-power-law exponent $-4.5<\ss<8.2$.}
    \label{fig:enter-label}
\end{figure}

\begin{figure}[h]
    \centering
    \includegraphics[scale=1]{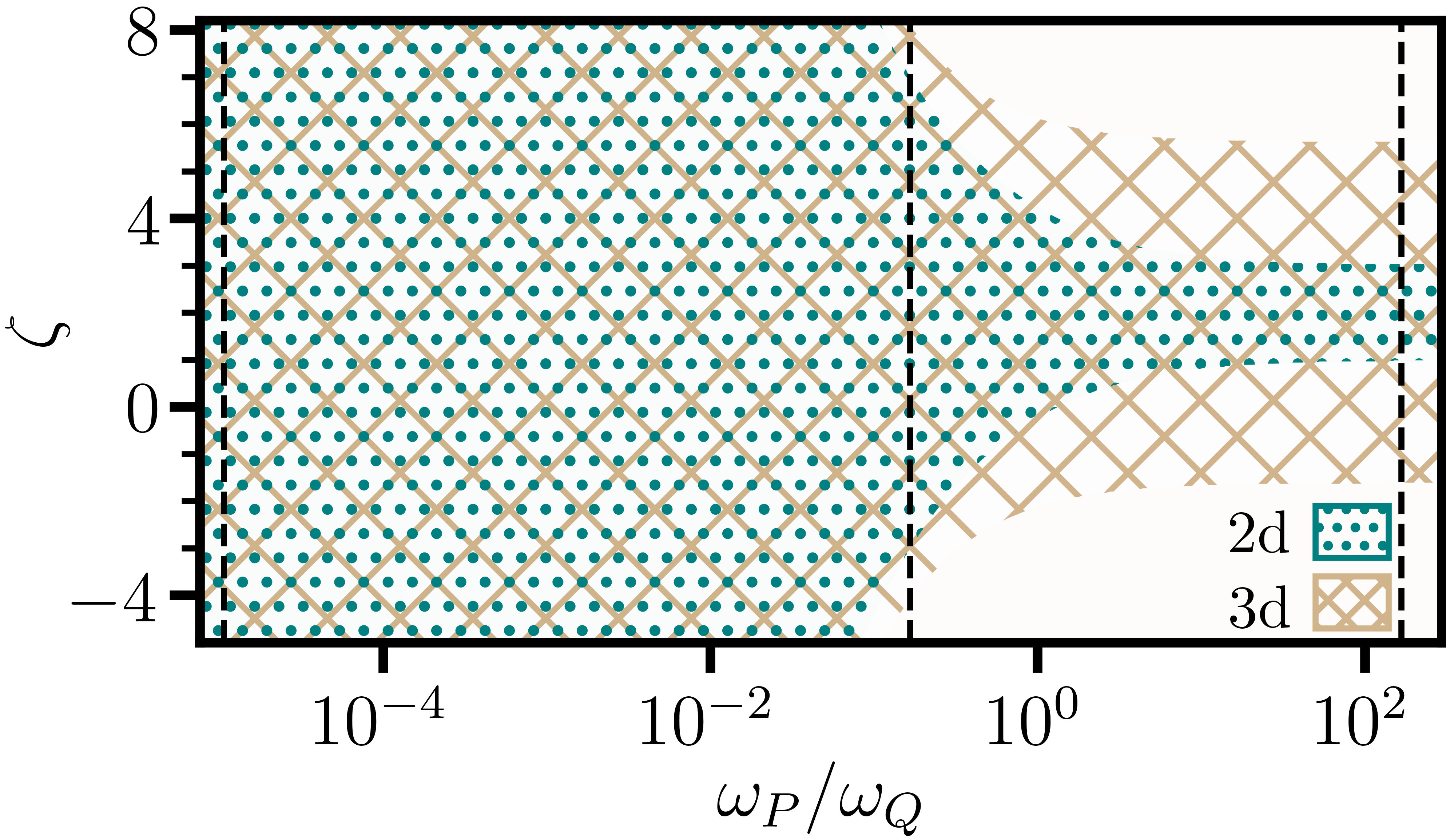}
    \caption{The area in ($\wp/\wq-\zeta$) plane where eccentric modes could exist for 2D (dark green dotted area) and 3D (brown hatched area) locally isothermal disks, plotted using Eq. (\ref{eq:condition on disk parameters for modes}). The vertical black dashed lines mark the three reference $\wp/\wq$ values from Table \ref{table:cav_ratio}.}
    \label{fig: area in zeta-wp/wq space where modes are possible}
\end{figure}

In Figure \ref{fig:enter-label}, we show how the magnitude and the sign of the pressure potential changes with the power-law-density exponent $\ss$. For both 2D and 3D locally isothermal disks, for intermediate $\ss$ values given above, the pressure potential is positive. For those values of $\zeta$, the condition given by Eq. (\ref{eq:condition on disk parameters for modes}) is satisfied for any value of $\wp/\wq$, and trapped modes are always possible. But, for large $|\ss|$, the pressure potential is negative, and the condition given by Eq. (\ref{eq:condition on disk parameters for modes}) is satisfied only for some values of $\wp/\wq$. To illustrate this, in Figure \ref{fig: area in zeta-wp/wq space where modes are possible}, we mark the area in $(\wp/\wq,\zeta)$ plane for which trapped modes are possible (for which Eq. (\ref{eq:condition on disk parameters for modes}) is satisfied). We see that for larger values of $|\zeta|$, there is a maximum value of $\wp/\wq$ for which trapped modes are possible. But for intermediate values of $\zeta$, equal to those for which $-(\nu^2-1/4)$ is positive in Figure \ref{fig:enter-label}, trapped modes are possible for any value of $\wp/\wq$.

\begin{figure}[h]
    \centering
    \includegraphics[scale=1]{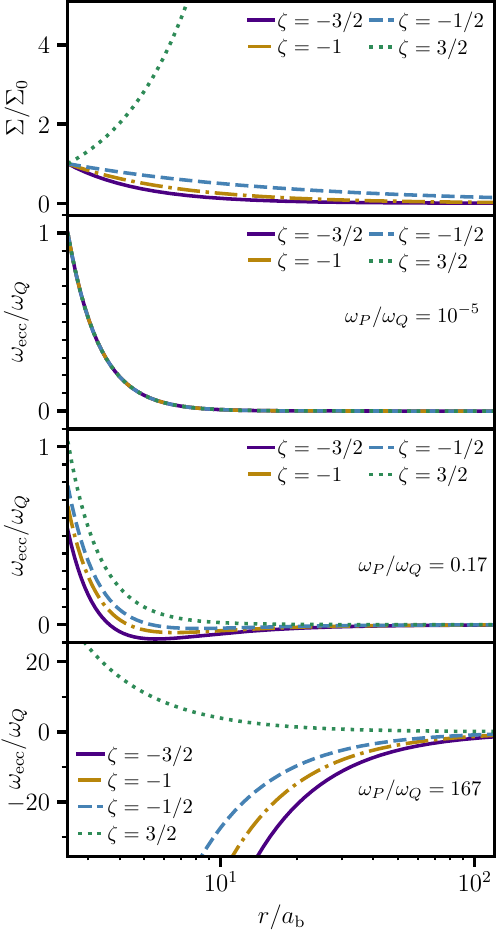}
    \caption{Density profiles $\Sigma=\Sigma_0(r/\rin)^{\ss}$ (upper panel), and the total eccentricity potential for the three reference values of the pressure to gravity frequency ratio $\wp/\wq$ given in Table \ref{table:cav_ratio}; $\wp/\wq=10^{-5}$ (second panel), $\wp/\wq=0.17$ (third panel), and $\wp/\wq=167$ (bottom panel).}
    \label{fig:power-law potentials}
\end{figure}

In Figure \ref{fig:power-law potentials}, we plot examples of density profiles and corresponding eccentricity potentials for 2D locally isothermal disks for four values of the power-law-density exponent $\zeta=\{-3/2,-1,-1/2,3/2\}$ for the three reference values of the pressure-to-gravity frequency ratio $\wp/\wq$ given in Table \ref{table:cav_ratio}. For thin disks, \ie small $\wp/\wq$ values, the total eccentricity potential is same for all density profiles, and equal to $\omecc\approx\omg$ (second panel of Figure \ref{fig:power-law potentials}). For intermediate $\wp/\wq$ values, the eccentricity potentials for different density profiles differ, but are all positive near the disk inner edge (third panel of Figure \ref{fig:power-law potentials}). For thick disks, \ie high $\wp/\wq$ values (bottom panel of Figure \ref{fig:power-law potentials}), the eccentricity potentials are negative for $\ss=\{-3/2,-1,-1/2\}$, due to a negative pressure eccentricity potential, whose magnitude is greater than the gravitational eccentricity potential. As a result, trapped modes are not possible. For $\ss=3/2$, the eccentricity potential is positive even for large $\wp/\wq$, due to a positive pressure eccentricity potential, so trapped modes could be possible.

\subsection{Non-precessing Modes}\label{sec:3a}

Eccentric, non-precessing circumbinary disks have been reported as quasi-steady-state results of numerical hydrodynamical simulations \citep[\eg,][]{Miranda_Munoz_Lai_2017, Dan-transition, Siwek+_2023}.
As a first step in understanding the spectrum of trapped modes in power-law disks, we seek to understand the eccentricity profile, and the conditions for the existence, of non-trivial non-precessing mode ($\omega=0$). 
We note that \cite{Lubow2022} demonstrates the existence of non-precessing free-modes of the eccentricity equation in connection to the resonant eccentricity growth that can arise in the presence of external forcing.

The eccentricity equation (Eq. (\ref{eq: eccentricity equation in terms of f_p and f_g})) for a non-precessing mode in a disk with power-law density profile as given by Eq. (\ref{eq: power-law density profile}) reduces to a Bessel equation. Its solutions $E_1(r)$ and $E_2(r)$ are products of a power of radius $r^{\delta}$, and Bessel functions of the first kind $J_{ \nu}$ and $J_{-\nu}$ (or second kind $Y_{-\nu}$ if $\nu$ is an integer). In Appendix \ref{appB - subsection: Bessel parameters}, we provide detailed calculations. The eccentricity profile (Eq. (\ref{eq: eccentricity profile solution in its E=b_1E_1+b_2E_2 form})) is then given by:
\begin{equation}\label{eq: non-precessing disks - full analytic solution for E(r)}
E(r)= \left(\frac{\rs}{r}\right)^{\delta} \left[\beta_1  J_\nu\left(\frac{\rs}{r} \right)+\beta_2 J_{-\nu}\left(\frac{\rs}{r} \right)\right]\,.
\end{equation}
Here, the constants $\beta_1$ and $\beta_2$ are determined by the boundary condition Eq. (\ref{eq: solutions for constants beta_1 and beta_2 in terms of E_1 and E_2}), the values of $\nu$ are given by Eq. (\ref{eq: nu Bessel parameter}), and the exponent $\delta$ is related to the power-law density exponent $\ss$ and the thermodynamic character of the perturbations via $\xi$ as follows
\begin{equation}\label{eq: delta Bessel parameter}
    \delta=-\frac{\ss}{2}-1+\frac{\xi}{2}.
\end{equation}
The radial scale in the eccentricity profile solution (\ref{eq: non-precessing disks - full analytic solution for E(r)}),
\begin{equation}\label{eq: rs Bessel parameter}
    \rs^2=\frac{\ab^2 Q}{h_e^2},
\end{equation}
sets the distance at which gravitational and pressure forcing are comparable, \ie, $\omg \sim \omp$; comparing equations Eq. (\ref{eq:omega_PQ}) and Eq. (\ref{eq: rs Bessel parameter}), we can write:
\begin{equation}\label{eq: scale radius in terms of frEq. ratio}
    \frac{\rs}{\rin} = \left(\frac{\wq}{ \wp}\right)^{1/2}.
\end{equation}

\subsubsection{Schrödinger's Insight on the Eccentricity Profile}
\label{subsec: Schrödinger's Insight on Eccentricity - POWER LAW}

The set of Eq. ((\ref{eq: nu Bessel parameter}),(\ref{eq: non-precessing disks - full analytic solution for E(r)})-(\ref{eq: delta Bessel parameter})) provides a full analytical solution for the structure of the non-precessing eccentricity mode. Valuable insight into the conditions for the existence of these solutions and their properties can be obtained by analyzing them in terms of Schrödinger's vernacular.

For a non-precessing mode, we can write the wavenumber defined in Eq. (\ref{eq: wave number k^2 for Schrödinger equation}) in terms of the Bessel function parameters given by Eqs. (\ref{eq: rs Bessel parameter})-(\ref{eq: nu Bessel parameter}) (see also Eq. (\ref{eq: k-nu relationship})) as
\begin{equation}
\label{eq: eccentricity potential in term of Bessel parameters}
       k^2=\rs^2r^{-4}-\left(\nu^2-\frac{1}{4} \right)r^{-2}.
\end{equation}
 
The radius $\rt$ at which the transition between radial oscillations to a steady decay of the eccentricity is found by setting $k^2(\rt)=0$, which implies
\begin{equation}
\label{eq: transition radius (non-precessing modes)}
\rt^2=\rs^2 \left(\nu^2-\frac{1}{4} \right)^{-1}.
\end{equation}
Thus, if $\nu^2<1/4$, $\rt$ has no real solutions and  $\omega=0$ can not be a localized state (trapped mode). This agrees with our previous conclusion that $\nu^2>1/4$ for the existence of trapped modes. Provided that $\nu^2>1/4$, whether or not non-trivial values of $\beta_1$ and $\beta_2$ satisfy the boundary condition will set constraints on the values of disk-binary model parameters ($Q,h$) that can host non-precessing trapped modes. We discuss this in more detail in \S \ref{subsection:Approximate solutions}.

\subsubsection{Spatial Behavior of Eccentric Modes}\label{subsection:Approximate solutions}
To characterize the properties of the solutions given by Eq. (\ref{eq: non-precessing disks - full analytic solution for E(r)}), and to find and simplify the condition for their existence, we use asymptotic forms for small and large values of the arguments of the Bessel functions \citep{Watson_1922}. Assuming a large outer radius\footnote{Note that $\beta_2\approx 0$ for a large enough outer radius. This can be seen by substituting $E_{1}=(r/\rs)^\delta J_{\nu} \left(\rs/r \right)$ and $E_{2}=(r/\rs)^\delta J_{-\nu} \left(\rs/r \right)$ (Eq. (\ref{eq: non-precessing disks - full analytic solution for E(r)})) into Eq. (\ref{eq: non-trivial constants beta_1 and beta_2  in terms of E_1 and E_2}) and taking $\lim_{\rout \rightarrow \infty} \beta_2/\beta_1=0$.}, the eccentricity profile presents an oscillatory and a non-oscillatory part, 
\begin{eqnarray} 
\label{eq: approximate E(r) for non-precessing disks}
    E(r)\propto\begin{cases}
     \left(r/\rs\right)^{\delta+1/2}\cos{\left( \frac{\rs}{r}-\frac{\nu\pi}{2}-\frac{\pi}{4}\right) };& r \ll \rt\\
      \left(r/\rs\right)^{\delta-\nu};& r \gg \rt
    \end{cases}.
\end{eqnarray}
The solutions for the radial eccentricity profile transition from oscillating, when gravity dominates, to decaying, when the negative pressure potential dominates over the positive gravitational eccentricity potential. 

For $r \ll \rt$, when gravitational forces dominate, the radial eccentricity oscillations are bounded by an envelope given by Eq. (\ref{eq: approximate E(r) for non-precessing disks}) as $\pm r^{\delta+1/2}$. We use Eq. (\ref{eq: delta Bessel parameter}) to write this envelope in terms of the power-law-density exponent $\zeta$ as $\pm r^{\delta+1/2}=r^{(\ss+\xi-1)/2}$. If $\ss<1-\xi$, the amplitude of oscillations gets smaller as the radius increases, so the maximum value of the eccentricity always occurs at the disk inner edge. By Eq. (\ref{eq: delta Bessel parameter}), such decreasing eccentricity profiles, with $\delta+1/2<0$, are possible only for $\xi=0$ and $\ss> 1$. Note this is consistent with the results obtained for the values chosen in \cite{ML2020}, where a main conclusion is that the disk eccentricity profile is always largest at the inner edge. We note, however, that decreasing eccentricity profiles are not the only allowed class of solutions. If instead, $\delta+1/2>0$, as for example when $(\xi=0,\ss>-1/2)$ or $(\xi=1, \ss>0)$, then the amplitude of oscillations increases for larger radii, so the maximum value of disk eccentricity must occur at $\rt$ (Eq. (\ref{eq: transition radius (non-precessing modes)})). 
We further explore eccentricity profiles and their envelope of oscillations in \S\ref{subsection:Examples} below.

\subsubsection{The Inner Gravity-to-Pressure Ratio as Control Parameter}

We find (approximate) conditions for the existence of non-precessing modes by substituting the asymptotic expressions for the analytical solutions, Eqs. (\ref{eq: approximate E(r) for non-precessing disks}), into the boundary condition 
(Eq. (\ref{eq: the boundary condition - general})), using the result (\ref{eq: scale radius in terms of frEq. ratio}), and approximating the condition for the existence of non-trivial solutions $\Psi_{\rm BC}=0$ as (Eqs. \ref{eq: solutions for constants beta_1 and beta_2 in terms of E_1 and E_2} and  \ref{eq: zeta definition}): 
\begin{equation}\label{eq: zeta before approx}
    \left(\delta+\frac{3}{2}-\xi \right)\sqrt{\frac{\wp}{\wq}}=\tan{\left(\sqrt{\frac{\wq}{\wp}}-\frac{\nu\pi}{2}-\frac{\pi}{4} \right)}.
\end{equation}
For $\wp/\wq\ll 1$, the solution to Eq. (\ref{eq: zeta before approx}) is:
\begin{equation}
\label{eq: zeta approximation (non-precessing power law disks)}
    \Psi_{\rm BC} \approx \sqrt{\frac{\wq}{\wp}}-\frac{\nu \pi}{2}-\left(n+\frac{1}{4} \right)\pi=0,
\end{equation}
where $n\in \mathbb{N}_0$ is the order of the non-precessing eccentricity mode. This shows explicitly that the argument of the cosine in Eq. (\ref{eq: approximate E(r) for non-precessing disks}) depends on $n$, showing that the mode order sets the number of zeros of $E(r)$. In particular, we refer to the $n=0$ mode with no nodes as the ground mode.

Let us note that the mode number $n$ in Eq. (\ref{eq: zeta approximation (non-precessing power law disks)}) and the transition radius $\rt$ (Eq. (\ref{eq: transition radius (non-precessing modes)})) are increasing functions of the binary gravitational parameter $Q$, and decreasing functions of the disk aspect ratio $h$. That is, for fixed disk inner radius, density profile $\ss$, and adiabatic coefficient $\gamma$, a higher mode number $n$, corresponds to solutions that exist only for larger values of $(\wp/\wq) \propto h^2/Q$. With that in mind, comparing Eq. (\ref{eq: scale radius in terms of frEq. ratio}) to Eq. (\ref{eq: transition radius (non-precessing modes)}), we conclude that higher-order modes span larger disk radii (see Eq. (\ref{eq: rs Bessel parameter})).

As a corollary of these findings, we conclude that, for power-law disks, the ratio $\rs/\rin = \sqrt{\wq/\wp}$ uniquely determines the existence and properties of non-trivial, non-precessing modes.

\subsubsection{Implications for Representative Disk Model}
\label{subsection:Examples}
To close this section on analytical characterization of non-precessing modes, we illustrate the radial extent of oscillations, the amplitude of these oscillations, and the mode order of zero-precession disk eccentricity with three examples. 

\begin{figure}[t!]
\includegraphics[scale=0.99]{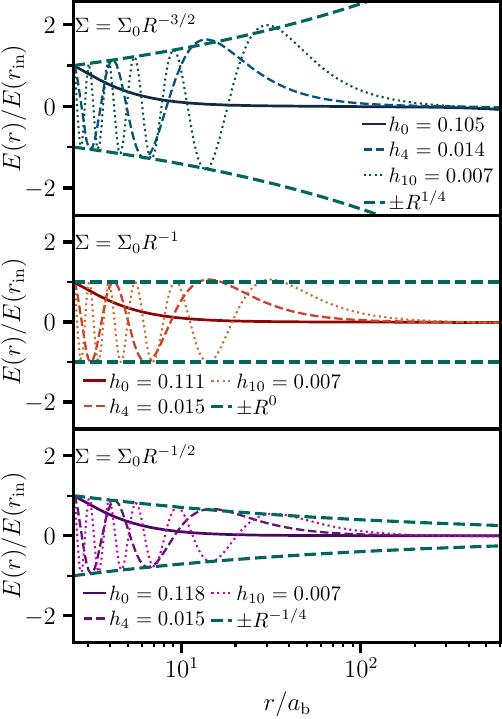}
\caption{Radial eccentricity profiles $E(r)$ for a 2D locally isothermal disk for three density profiles: $\Sigma \propto R^{-3/2}$ (upper panel, blue lines), $\Sigma \propto R^{-1}$ (middle panel, brown lines), and $\Sigma \propto R^{-1/2}$ (bottom panel, purple lines). In all three panels, we plot solutions for the ground, $n=0$ mode (solid lines), $n=4$ mode (dashed lines), and $n=10$ mode (dotted lines). In all panels, we plot eccentricity mode radial oscillation envelopes (thick-dashed dark green lines). The inner and the outer radii are $\rin=2.5\ab$ and $\rout=600\ab$, the binary is on a circular orbit with a binary mass ratio $q=0.5$. }
\label{fig: eccentricity profiles for non-precessing disks - 3 panels}
\end{figure}

Each panel in Figure \ref{fig: eccentricity profiles for non-precessing disks - 3 panels} illustrates the full analytical solutions for zero-precession eccentricity profiles, $E(r)$, for locally isothermal ($\xi=0$) 2D disks, for different values of the disk aspect ratio $h$, considered to be independent of $r$, and different values of the power-law-density exponent, $\ss=\{-3/2, -1, -1/2\}$, as shown in the top, middle and bottom panels, respectively. Each panel shows eccentricity profiles for three different mode orders $n=\{0, 4, 10\}$ (solid, dashed, and dotted, respectively). Also plotted are the oscillation envelopes as predicted by the prefactor in the first of the approximate solutions in Eq. (\ref{eq: approximate E(r) for non-precessing disks}) (green thick-dashed lines). 

To find solutions at a given mode order $n$, we find the values of the disk aspect ratio $h$ that satisfy the boundary condition (Eq. (\ref{eq: solutions for constants beta_1 and beta_2 in terms of E_1 and E_2})) for that $n$. For these values of $h$ (labeled in each panel), we calculate constants $\beta_1$ and $\beta_2$ given by Eq. (\ref{eq: non-trivial constants beta_1 and beta_2  in terms of E_1 and E_2}) to plot eccentricity profiles given by Eqs. (\ref{eq: non-precessing disks - full analytic solution for E(r)}-\ref{eq: nu Bessel parameter}). We scale all solutions with their values at the disk's inner edge. 

For each of the three density profiles, smaller values of $h$ correspond to higher-order modes, as expected from the approximation given by Eq. (\ref{eq: zeta approximation (non-precessing power law disks)}). From this equation, we infer that the higher the mode order, the smaller effect the exact value of the power-law-density exponent $\ss$, through $\nu$ via Eq. (\ref{eq: nu Bessel parameter}), has on the solutions for the disk aspect ratio $h$.  As a result, out of the three plotted modes, it is only the value of $h_0$, corresponding to the ground-state mode, that changes for different density profiles. Notice that each of the ground state modes changes very little with the density profile. This is because the density profile does not affect the decaying portion of the solution in Eq. (\ref{eq: approximate E(r) for non-precessing disks}). Specifically, using the expressions for $\delta$, in Eq. (\ref{eq: delta Bessel parameter}), and $\nu$, in Eq. (\ref{eq: nu Bessel parameter}), we find that the decaying part of the eccentricity profile for a 2D, locally isothermal power-law disk is given by $r^{\delta-\nu}=r^{(\ss/2-1)-(\ss/2+1)}=r^{-2}$ for any $\ss$.

The density profile strongly affects the envelope of oscillations. From Eq. (\ref{eq: nu Bessel parameter}) and the first of Eqs.~(\ref{eq: approximate E(r) for non-precessing disks}), and defining $R=r/\ab$, the oscillation envelope for the examples in Figure \ref{fig: eccentricity profiles for non-precessing disks - 3 panels} reduce to $\pm R^{(\ss-1)/2} = \pm \{R^{1/4}, R^{0}, R^{-1/4}\}$, for top, middle, and bottom panels, respectively. Consequently, for $\ss=1$ (middle panel), the amplitudes of the cosine oscillations are scale-free. For density profiles shallower than $\Sigma\propto R^{-1}$, the maximum value of the eccentricity is always at the disk's inner edge (bottom panel). For density profiles steeper than $\Sigma\propto R^{-1}$, the maximum value of the eccentricity can be found near the transition radius $\rt$ (upper panel). Therefore, by Eq. (\ref{eq: transition radius (non-precessing modes)}), the radius at which the disk is most eccentric depends sensitively on the value of the disk aspect ratio. For all three density profiles and mode orders, $\pm R^{(\ss-1)/2}$ is an excellent fit for the envelope everywhere even at the transition radius where, in principle, $r/\rt \sim 1$ invalidates the condition necessary for the Bessel function approximation used in Eq. (\ref{eq: approximate E(r) for non-precessing disks}). 

We can characterize the behavior of the radial solutions using Eqs.~(\ref{eq: approximate E(r) for non-precessing disks}) and (\ref{eq: transition radius (non-precessing modes)}). If the inner boundary is larger than the transition radius, $\rin \gtrsim \rt$, $E(r)$ exhibits no radial oscillations and its maximum value is located at $r=\rin$. But, if $\rin< \rt$, the location of the eccentricity maximum depends on the details of the density profile and the thermodynamic properties of the perturbations ($\xi$).

\subsection{The Number of Modes}\label{sec:3b}
We extend the analytic treatment of eccentricity modes in disks with a constant disk aspect ratio and a power-law-density profile to precessing modes.
In Section \ref{sec:3a}, we showed how the (free) mode order $n$ changes with the disk and binary parameters if we focused on the non-precessing ($\omega=0$) mode. 
For example, in Figure \ref{fig: eccentricity profiles for non-precessing disks - 3 panels}, we saw that thinner non-precessing disks corresponded to higher-order modes, as was shown in \cite{Lubow2022}. Similarly, works like \cite{ML2020} show that multiple precessing modes (i.e., $\omega\ne0$) can exist for thinner disks. As an application of the results derived in Section \ref{sec:3a}, we quantify the number of allowed modes as a function of binary and disk parameters.

First, we find an analytical estimate for the number of modes (\S\ref{sec:modes}), and then we find all $\omega_n$ solutions for two illustrative examples (\S \ref{subsection: precessing modes - Implications for a Representative Disk Model}).

\subsubsection{Analytical Approximation}\label{sec:modes}

We find an analytical description of the number of different eccentricity modes $N$ that can occur in a given disk-binary system. That is, we find the number of different values of $\omega$ for which, with boundary conditions imposed, a non-trivial eccentricity profile $E(r)$ exists. 

\begin{figure}[t!]
    \centering
    \includegraphics[scale=1.0]{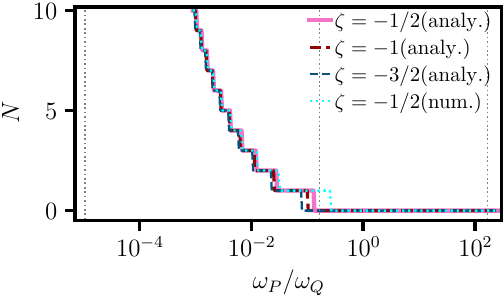}
    \caption{The number of possible eccentricity modes $N$ as a function of the binary and disk parameters for 2D locally isothermal disks with power-law-density profiles for three values of the power-law exponent $\ss$. We plot solutions given by Eq. (\ref{eq: number of modes}) for $\ss=-1/2$ (light pink solid line), $\ss=-1$ (red dashed line), and $\ss=-3/2$ (blue dashed line). The solutions are not sensitive to the value of $\ss$ over a wide range of feasible disk and binary parameters. We plot numerical solutions for $\ss=-1/2$ (dotted cyan purple line). The analytic estimate given by Eq. (\ref{eq: number of modes}) is an excellent match for all but the lowest mode. The vertical gray dotted lines mark the three reference $\wp/\wq$ values from Table \ref{table:cav_ratio}.}
    \label{fig: the number of modes N for power law disks}
\end{figure}

In the absence of analytic solutions for the eccentricity of precessing disks, we find approximate values of $N$ without providing solutions for the frequency of the $n$-th mode $\omega_n$. To do so, we use the following reasoning. First, we note that the lowest trapped state frequency is $\omega \gtrapprox 0$ (Eq. (\ref{eq: the range of trapped mode frequencies in terms of Bessel parameters - v2})). Second, for a given eccentricity potential, the mode order $n$ grows as the value of its eigenvalue $\omega_n$ decreases\footnote{We can see this from the WKB quantization condition, which states $\int_{r_\mathrm{1}}^{r_{\mathrm{2}}}k(r)dr=\pi/2+n\pi$, where $n$ is the mode order and $r_\mathrm{1}$ and $r_\mathrm{2}$ are the turning points for which $k^2=0$. For a given binary-disk system (fixed eccentricity potential), $k^2$ increases with increasing $\omega$ (Eq. (\ref{eq: wave number k^2 for Schrödinger equation})).}. Consequently, the lowest-frequency solution corresponds to the highest-order mode. We conclude that we can find the number of modes a disk can support by finding the mode order of its lowest frequency solution, which we argued is $\omega \geq 0$. In \S\ref{sec:3} (Eq. (\ref{eq: zeta approximation (non-precessing power law disks)})), we found exactly that; the mode order $n$ for the lowest frequency mode $(\omega=0)$ for binary-disk systems that allow for a non-precessing mode. We use that result (Eq. (\ref{eq: zeta approximation (non-precessing power law disks)})) to approximate the number $N$ of possible solutions $\omega_n$ for other system parameters for which we know neither the frequency nor the order of the highest order mode:
\begin{equation}
\label{eq: number of modes}
    N\approx \left\lfloor \frac{\sqrt{Q}}{\pi h_e \Rin}-\frac{\nu}{2}+\frac{3}{4}\right\rfloor =\left\lfloor\sqrt{\frac{\wq}{\wp}}\frac{1}{\pi} -\frac{\nu}{2}+\frac{3}{4}\right\rfloor,
\end{equation}
where we have accounted for the $n=0$ mode by setting $N=n+1$. The approximation given by Eq. (\ref{eq: number of modes}) is better for higher-order modes because Eq. (\ref{eq: zeta approximation (non-precessing power law disks)}) was derived under the assumption $\rs/\rin\gg1$, which is true for large $n$ (as seen from Eq. (\ref{eq: zeta approximation (non-precessing power law disks)})). We compare the approximation given by Eq. (\ref{eq: number of modes}) with numerical results next, and later in \S\ref{subsection: Disk Density Impact on Eccentricity - The number of modes}.

For better approximation of $(\wp/\wq, \nu)$ values that allow for one mode ($N=1$), the alternative to Eq. (\ref{eq: number of modes}) is Eq. (\ref{eq:condition on disk parameters for modes}). We discuss this more and provide an example in \S \ref{subsection: No Solutions for 2D Power-law Disks with high wp_wq ratio}.

In Figure \ref{fig: the number of modes N for power law disks}, we plot $N$ for a 2D locally isothermal disk, as a function of $\wp/\wq$, for different values of the density power-law exponent $\ss$ (solid lines). As a reference, we draw vertical dotted lines at three values of $\wp/\wq=h^2\Rin^2/Q$ from Table \ref{table:cav_ratio}. To test the validity of our analytical approximation across parameter space we also plot full numerical solutions (cyan dotted line) for the number of modes in Figure \ref{fig: the number of modes N for power law disks}. Analytical approximation of Eq. (\ref{eq: number of modes}) in Figure \ref{fig: the number of modes N for power law disks} show that the steepness of the density profile will influence the number of modes more for smaller values of $\wp/\wq$.  

\begin{figure}[t!]
    \centering
\includegraphics[scale=1]{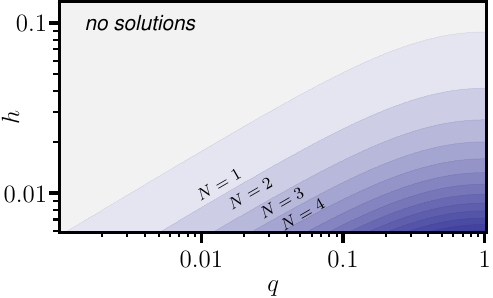}
    \caption{
    The number of possible eccentricity modes $N$ given by Eq. (\ref{eq: number of modes}) as a function of the binary mass ratio $q$ and the disk aspect ratio $h$. The disk is assumed to be a 2D locally isothermal disk with a density profile $\Sigma\propto R^{-1/2}$, and an inner radius $\rin=2.5\ab$. In light gray, we mark the area for which no solutions exist. In blue, we draw selected contours of constant $N$.  Lines of constant $\wq/\wp$ are parallel to the drawn contours. Lines dividing contoured areas correspond to $(q,h)$ values for which a non-precessing ($\omega=0$) mode is possible. 
}
    \label{fig: number of modes - color coded}
\end{figure}

\subsubsection{Implications for a Representative Disk Model}\label{subsection: precessing modes - Implications for a Representative Disk Model}

By Eq. (\ref{eq: number of modes}) and Eq. (\ref{eq:omega_PQ}) , the number of possible eccentricity modes increases with increasing binary mass ratio, and decreases with increasing disk aspect ratio. 
This is shown in Figure \ref{fig: number of modes - color coded}, where we plot the number of modes given by Eq. (\ref{eq: number of modes}) as a function of the binary mass ratio $q$ and the disk aspect ratio $h$ in a 2D locally isothermal disk with density profile $\Sigma\propto r^{-1/2}$ and an inner radius of $\rin=2.5\ab$, around a binary on a circular orbit. 

In Figure \ref{fig: number of modes - color coded}, each contour level represents the region of parameter space where the combination of $(q,h)$ values results in $N$ possible modes. In the gray region above the $N=1$ contour level, no trapped modes exist.  
For a circumbinary disk characterized by a pair of $(q,h)$ values on the line delimiting the regions denoted by $N$ and $N+1$, with $N\geq 0$, there exist $N+1$ different solutions. On this line, the lowest frequency mode is non-precessing and all other frequency solutions are real and positive, with\footnote{For $N=0$ the ground mode is non-precessing, i.e., $\omega_{N}=\omega_0=0$, which can happen only on the line delimiting the no-solution region.} $0=\omega_{N} < \omega_{k} < \omega_{0}$, with $k=1,\ldots, N-1$. Similarly, for systems with $(q,h)$ values on the contour level labeled by $N+1$ in Figure \ref{fig: number of modes - color coded}, $N+1$ different solutions exist, all with real positive precession frequencies.

Figure \ref{fig: number of modes - color coded} shows that as we consider lower values of $h$, thinner disks, and larger values of $q$, towards equal-mass binaries, the number of modes increases steeply.

\begin{figure}
    \centering
    \includegraphics[scale=0.98]{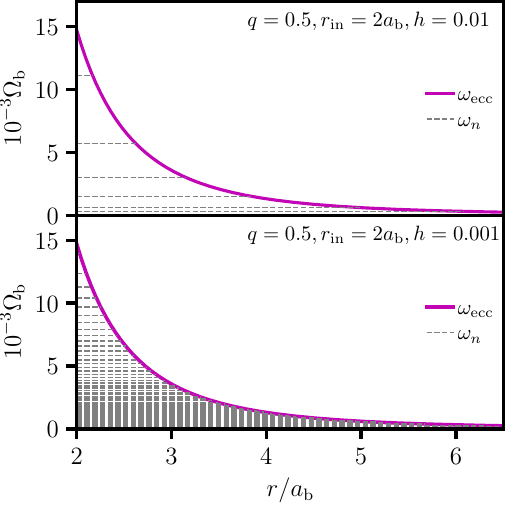}
    \caption{The eccentricity potential (solid purple lines) and values of precession frequencies corresponding to each of the possible modes (dashed grey lines) in 2D disks with $h=10^{-2}$ (top panel) and $h=10^{-3}$ (bottom panel).    }
    \label{fig: the eccentricity potential + omega solution spectrum for power law disks}
\end{figure}

For thin disks, decreasing the value of the disk aspect ratio $h$ increases the number of possible modes (Figure \ref{fig: number of modes - color coded}). It also decreases the spacing of the mode frequencies. This happens because, for $\wp/\wq\ll 1$, the eccentricity potential is independent of $\wp$ as $\omecc\approx \wq (\rin/r)^{7/2}$ (Eq. (\ref{eq: ecc pot power})), but the number of modes is inversely proportional to the disk aspect ratio with $N\approx \sqrt{\wq/\wp}\propto 1/h$ (Eq. (\ref{eq: number of modes})). To illustrate this, in Figure \ref{fig: the eccentricity potential + omega solution spectrum for power law disks}, we plot eccentricity potentials (solid lines) and all possible $\omega_n$ values (dashed lines) for $q=0.5$ and  $h=0.01$ ($\wp/\wq=1.2\times 10^{-3}$) (upper panel) and $h=0.001$ ($\wp/\wq=1.2\times 10^{-5}$) (bottom panel). In Figure \ref{fig: the eccentricity potential + omega solution spectrum for power law disks}, the eccentricity potential exhibits no visible changes when decreasing the value of $h$ ten-fold, but the number of modes increases significantly, leading to more densely spaced mode frequency values.

\section{Disk Density Impact on Eccentricity} \label{sec: Disk Density Impact on Eccentricity}

For any application of the formalism presented in Section \ref{sec:methods}, we must choose the circumbinary disk's radial density profile. Here we move beyond the power-law density profile of the previous sections to show how the choice of a more realistic set of density profiles, applicable to CBDs, influences the ground-mode precession frequency $\wo$, the number of possible eccentricity modes $N$, the eccentricity potential $\omecc$, and the eccentricity profile $E(r)$. We identify properties insensitive to this choice to show the applicability of analytic solutions presented in Section \ref{sec:3a} and Section \ref{sec:3b} to a wider range of disk density models. 

Understanding the sensitivity of results to the choice of the density profile also allows us to gauge when the presented eccentricity formalism can be reliably applied to interpret numerical hydrodynamical simulations or to make predictions for not-yet-simulated parts of parameter space (assuming linear theory can be applied). To determine the sensitivity of the solutions to the choice of the disk density profile, we solve the eccentricity boundary value problem numerically for three different classes of density-profile models.

\subsection{Three Useful Density Models} \label{subsection: Disk Density Impact on Eccentricity - Three Useful Density Models}

{\it Power-law disk} \/-- We first consider a density profile corresponding to a standard $\alpha$-disk \citep{Shakura_Sunyaev_1973} with constant $\alpha$ and constant aspect ratio $h$. In such a case, the coefficient of (turbulent) kinematic viscosity $\nu =\alpha c_{\rm s} H$, must be $\nu\propto R^{1/2}$, with the steady-state disk surface density
\begin{equation}\label{eq: sigma_1}
\Sigma_1(R)=\Sigma_0 R^{-1/2},
\end{equation}
where $\Sigma_0$ is a constant that sets the global density scale and $R \equiv r/\ab$. This density profile corresponds to the power-law disk with $\ss =-1/2$ we already considered in \S \ref{sec:3}.

{\it Power-law disk with central cavity} \/-- The angular momentum exchange between the disk and the binary can modify the density profile given by Eq. (\ref{eq: sigma_1}) near the binary. 
Given a disk with a specified aspect ratio and viscosity, the smallest mass ratio systems will not generate significant deviations to the background power-law-density profile. For larger mass ratios ($q\sim 10^{-3}$), the non-linear wake excited by the secondary can clear an annular gap in the disk \citep[\eg,][]{Ward97, Duffell_MacFadyen_2013, Dan-gaps}. For large enough mass ratios, ($q\gtrsim0.05$) the binary can clear a central cavity in the disk \citep{Artymowicz_Lubow_1994,Dorazio_Haiman_2013,Farris_Duffel_2014, Dan-gaps, NotLindblad, DittmannRyan_q:2023}. A density-profile model that accounts for these binary effects is \citep{Miranda_Munoz_Lai_2017,ML2020},
\begin{equation}\label{eq: sigma_2}
\Sigma_2(R)=\Sigma_0 R^{-1/2}\left(1-L_0 R^{-1/2} \right)e^{-(R_\mathrm{cav}/R)^Z},
\end{equation}
where $L_0\equiv l_0/(\ob \ab^2)$, $l_0$ is the net angular momentum per unit mass gained/lost by the binary, and the parameter $Z$ controls the steepness of the cavity of characteristic size $\rcav$. The value of $L_0$ and the characteristics of the cavity are determined by the system parameters such as the binary mass ratio and the disk viscosity 
\citep{Miranda_Munoz_Lai_2017,Moody_2019,Duffel_2020,Tiede_2020,Dan-transition, Dittmann_Ryan_2022, DittmannRyan_q:2023, TiedeDOrazio:2024}. However, in order to understand how sensitive the eccentricity profile is to the values of $L_0$, $\Rcav\equiv\rcav/\ab$, and $Z$, we vary these quantities independently. 
We consider the ranges
\begin{eqnarray}
    L_0 &=& \{-0.2, 0.7, 1.0 \} \,, \\
    \Rcav &=& \{2.0, 2.5, 3.5\} \,,  \\
    Z &=& \{9, 12, 24, 48\} \,.
\end{eqnarray}
We use $L_0=0.7$, $\Rcav=2.5$, and $Z=12$ as fiducial values to facilitate comparison with the results presented by \cite{ML2020}.

\begin{figure}[t!]
    \centering
    \includegraphics[scale=1]{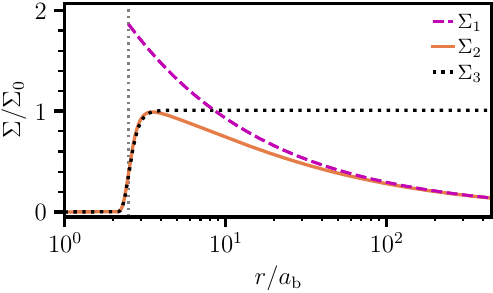}
    \caption{Radial density profiles given by Eq. (\ref{eq: sigma_1}) (dashed purple line), Eq. (\ref{eq: sigma_2}) (solid orange line), and Eq. (\ref{eq: sigma_3}) (dotted black line) for fiducial values $L_0=0.7$, $\Rcav=2.5$, and $Z=12$. The vertical gray dotted line marks the cavity radius. All three density profiles are scaled differently. }
    \label{fig: radial density profiles for the 3 density models}
\end{figure}

{\it Constant density disk with central cavity} \/-- Finally, to isolate the significant influence of the density profile beyond the central cavity shape, we consider a disk model with the same central cavity as the fiducial density profile $\Sigma_2$ but with uniform density beyond, i.e., 
\begin{equation}\label{eq: sigma_3}
    \Sigma_3(R)=\Sigma_0 e^{-(2.5/R)^{12}}.
\end{equation}

In Figure \ref{fig: radial density profiles for the 3 density models}, we plot the three density profiles, $\Sigma_1$, $\Sigma_2$, and $\Sigma_3$ for our chosen fiducial parameters. For the density profile $\Sigma_1$, the cavity size is effectively set by the inner disk radius $\rin$ discussed in Section \ref{sec:3}, so here we set it equal to the fiducial cavity radius, $\rin =2.5\ab$.

\subsection{2D Disk Precession for Different Density Profiles}
\label{subsection: Disk Density Impact on Eccentricity - The effect of a central cavity in 2D disks}

We solve the eccentricity boundary value problem numerically\footnote{Refer here to the appendix providing the details of how to choose $\rin$.}, as described in \S\ref{S:Methods_numerics}, to find the ground mode eccentricity precession frequency $\wo$ in 2D disk models with density profiles given by $\Sigma_1(r)$, $\Sigma_2(r)$, and $\Sigma_3(r)$. Figure \ref{fig: frequency solutions for 3 density models - 2D} shows the solutions $\omega_0$ normalized by the sum of the gravitational quadruple and pressure induced frequencies at the cavity radius, $\wq+\wp$. The left hand side of Figure \ref{fig: frequency solutions for 3 density models - 2D} is similar to $\wo/\wq$ solution plots from \cite{ML2020} because for $\wp/\wq\ll 1$, the y-axis values are simply $\wo/(\wq+\wp)\approx \wo/\wq$. An equivalent interpretation is possible for the right hand side of Figure \ref{fig: frequency solutions for 3 density models - 2D}, where, for $\wp/\wq\gg 1$, $\wo/(\wq+\wp)\approx \wo/\wp$. Each panel in this figure shows the solutions corresponding to $\Sigma_2(r)$, obtained by independently varying the values of $L_0$ (upper panel), $\Rcav$ (middle panel)\footnote{For reference, \cite{ML2020}, in their Figure 3, show the same solutions that we plot in the middle panel of Figure \ref{fig: frequency solutions for 3 density models - 2D} (orange lines), but for a narrower range $0.002\lessapprox \wq/\wp \lessapprox 200$, and the solutions are presented in units of $\wq$.}, and $Z$ (bottom panel). Also plotted are solutions using $\Sigma_1(r)$ and $\Sigma_3(r)$ for comparison.
Several interesting features are revealed.

\subsubsection{Cavity Shape Determines the Ground Mode}
\label{subsubsection: Disk Density Impact on Eccentricity - Insights from Systematic Study of Fundamental Mode: Importance of Cavity Steepness}

\begin{figure}[t!] 
    \centering
    \includegraphics[scale=1]{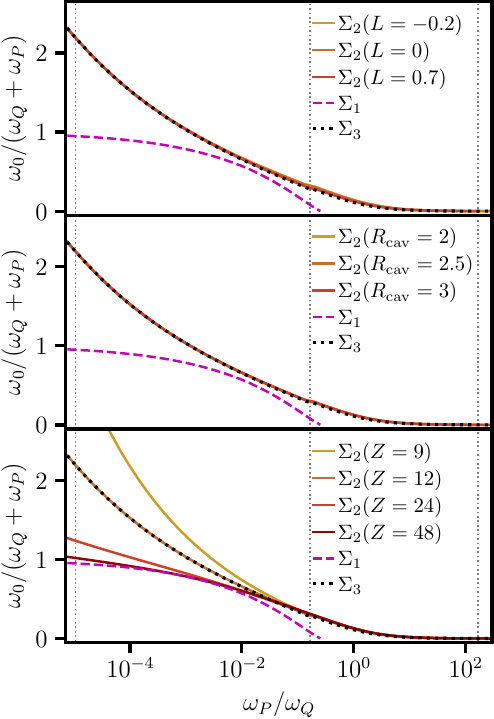}
    \caption{The ground mode precession frequency $\wo$ for 2D locally isothermal disks with radial density profiles given by Eq. (\ref{eq: sigma_1}) (dashed purple line), Eq. (\ref{eq: sigma_2}) (solid orange lines) and Eq. (\ref{eq: sigma_3}) (dotted black line). The fiducial parameters for $\Sigma_2$ are $L_0=0.7$, $\Rcav=2.5$, and $Z=12$. We vary the values of $\Sigma_2$ parameters $L_0$ (top panel), $\Rcav$ (middle panel), and $Z$ (bottom panel). The vertical gray dotted lines mark the three reference $\wp/\wq$ values from Table \ref{table:cav_ratio}.  The leading parameter in determining the precession frequency of thin disks ($\wp/\wq\lessapprox 1$) is the steepness of the disk's central cavity. Thin disks with shallower cavities precess faster.     }
    \label{fig: frequency solutions for 3 density models - 2D}
\end{figure}

The top and middle panels of Figure \ref{fig: frequency solutions for 3 density models - 2D} show that for disks with relatively ``soft'' cavities, i.e., $Z\simeq 10$, the frequency of the ground mode is largely insensitive to either $L_0$ or $\rcav$. Note that this does not imply that $\wo$ is independent of the cavity size $\Rcav$, as this sets the scale of the quadrupole precession frequency. 
It is already clear from these two panels that the value of the ground mode precession frequency is determined by the shape (steepness and location) of the cavity but not to the density profile beyond it.

The bottom panel of Figure \ref{fig: frequency solutions for 3 density models - 2D} indeed confirms that it is the steepness of the cavity that solely controls the functional dependence of $\wo/(\wq+\wp)$ on $\wp/\wq$. There are two regimes depending on the value of $\wp/\wq$. For thin disks, $\wp/\wq \ll 1$, the steeper the cavity the lower the ground-mode precession frequency $\wo$. The lowest limit is reached for the sharpest cavities, which converge to the values obtained for the power-law disk density model. As expected, for $\wp/\wq \ll 1$, the precession frequency tends to the quadrupole frequency. Beyond $\wp/\wq \gtrsim 1$, all the density profiles with ``smooth'' cavities lead to the same, constant ground-mode precession frequency $\wo\approx0.2 \wq$ \citep{ML2020}. These solutions do not coincide with the one obtained for a ``sharp'' cavity as implied by setting the boundary conditions at $\rin=\rcav$ for a power-law disk density profile. In fact, for $\wp \gtrsim \wq$, disks with sharp cavities cannot host trapped modes (see below.)

In summary, our analysis shows that the existence of a central cavity, more specifically its shape (steepness) is crucial in determining the precession frequency of thin disks with $\wp/\wq \ll 1$, see Figure \ref{fig: frequency solutions for 3 density models - 2D}. Since the steepness of the cavity (controlled by the parameter $Z$) is the only factor governing $\wo/(\wq+\wp)$, in what remains of this section, we fix the values of $L_0$ and $\Rcav$ to fiducial ones.

Most numerical hydrodynamical simulations of circumbinary disks study systems with values of $\wp/\wq \simeq 0.1-1$ (see Table \ref{table:cav_ratio}), and a central cavity. Hence, the analysis surrounding Figure \ref{fig: frequency solutions for 3 density models - 2D} predicts that, in this intermediate regime, the disk precession frequency is not sensitive to the shape of the disk density profile, but rather it is set by the value of $\wp/\wq$ (given by the disk aspect ratio, binary parameters, and cavity radius) for a wide range of density profiles.

\subsubsection{Insights From the Eccentricity Potential}\label{subsubsection: Disk Density Impact on Eccentricity - The eccentricity potential as an explanation for the solution dependence on cavity steepness}

We can understand why the shape of the central cavity is the key factor determining the precession frequency of the fundamental mode $\wo$ for low values of $\wp/\wq$ (thin disks), and why softer cavities result in faster precession (Figure \ref{fig: frequency solutions for 3 density models - 2D}) by analyzing the behavior of the eccentricity potential.
\begin{figure}
    \centering
    \includegraphics[scale=1]{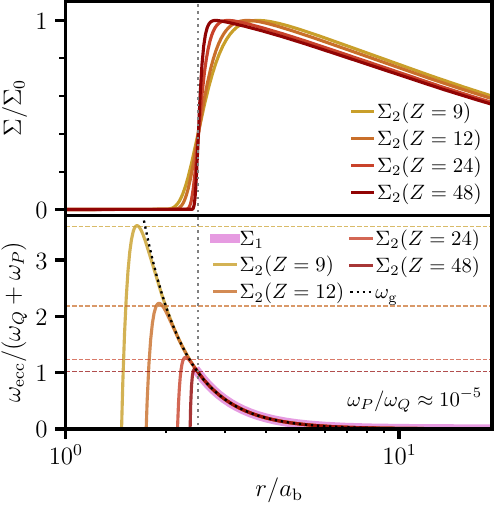}
    \caption{ Density profiles $\Sigma_2$ with a cavity size $\rcav=2.5\ab$ for four cavity steepness parameters $Z=\{9,12,24,48\}$ (upper panel) and their eccentricity potentials $\omecc$ in 2D locally isothermal disks for $\wp/\wq=10^{-5}$ (bottom panel). For all four eccentricity potentials, in the bottom panel, we plot the value of the ground mode precession frequency $\wo$ (thin dashed vertical lines). In the bottom panel, we also plot the gravitational eccentricity potential (dotted black line) and the eccentricity potential for $\Sigma_1$ (solid thick purple line). In both panels, the thin dotted vertical line denotes the cavity radius $\rcav$. Density profiles are scaled to the maximum value of $\Sigma/\Sigma_0=1$. Eccentricity potential shape and location changes significantly with changing the cavity steepness parameters $Z$. Ground mode precession frequencies are approximately equal to the maximum of the eccentricity potential.  }
    \label{fig: eccentricity potentials for different values of Z}
\end{figure}

In Figure \ref{fig: eccentricity potentials for different values of Z}, we plot the eccentricity potential $\omecc$ given by Eqs. (\ref{eq: eccentricity potential as a sum of its pressure and gravitational parts})-(\ref{eq: pressure part of the eccentricity potential}) for the same set of parameters as the bottom panel in Figure \ref{fig: frequency solutions for 3 density models - 2D} and for $\wp/\wq=10^{-3}$, where the solutions differ significantly for different values of the cavity steepness $Z$. A comparison of the solutions for the precession frequency (bottom panel of Figure \ref{fig: frequency solutions for 3 density models - 2D}) and the eccentricity potentials (Figure \ref{fig: eccentricity potentials for different values of Z}) shows that the precession frequency is approximately equal to the maximum of the eccentricity potential (horizontal dotted lines in the bottom panel). A comparison of the radial density profiles and the eccentricity potential in Figure \ref{fig: eccentricity potentials for different values of Z} suggests that the peak of the eccentricity potential ``well''  is located in the innermost parts of the disk, where the density is defined only by the (steepness of the) cavity shape (see Figure \ref{fig: radial density profiles for the 3 density models}). Since the frequency of a trapped mode should not be sensitive to the details of the shape of the potential outside of the potential ``well'', the localization of the mode near the cavity edge strongly suggests that knowing the shape of the cavity is enough to find the disk-precession frequency. 

\begin{figure}
    \centering
    \includegraphics[scale=0.95]{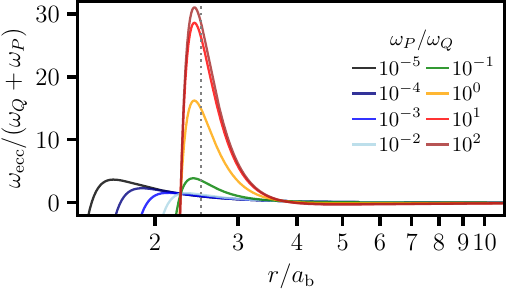}
    \caption{Eccentricity potentials $\omecc$ in units of $\wp +\wq$ for ten values of the cavity induced frequency ratios $\wp/\wq$ for the density profile $\Sigma_2$ with $Z=9$, $\rcav=2.5\ab$, and $L=0.7$.}
    \label{fig: eccentricity potentials for different values of wp/wq}
\end{figure}

In Figure \ref{fig: eccentricity potentials for different values of Z}, we also plot the gravitational part of the eccentricity potential $\omega_\mathrm{g}$ (dotted black line, Eq. (\ref{eq: gravitational part of the eccentricity potential})) to show why the ground mode frequency $\wo/\wq$ increases for softer cavities, \ie, for decreasing $Z$. All five eccentricity potentials are equal to the gravitational eccentricity potential beyond the peak of the potential well. The location of the potential well maxima shifts with changing the density profile for the same (low) value of $\wp/\wq$. Since the density profile information is contained only in the pressure eccentricity potential, we conclude that it is the pressure eccentricity potential that determines the location of the potential well for low values of $\wp/\wq$. In a disk with a soft cavity (low value of $Z$), a non-negligible amount of gas can be found near the binary, where the gravitational potential is deeper. Consequently, the gravitational eccentricity forcing increases with decreasing $Z$ because the disk spreads to lower radii. For very high values of $Z$, the cavity is so steep that it is similar to the sharp edge present in a power-law density profile. Because of this, the eccentricity potential becomes equal to that of a simple power-law-density disk (the purple and the dark red line in Figure \ref{fig: eccentricity potentials for different values of Z} overlap), resulting in the very similar precession frequency for both density profiles, as seen in Figure \ref{fig: frequency solutions for 3 density models - 2D}. 

In summary, for thin disks, $\wp\ll\wq$, pressure support is negligible and the peak of the effective potential extends significantly inside $\rcav$ provided there is enough gas there (the density profile of the cavity is ``soft''). In this case, the frequency of the ground mode is set by the quadrupole frequency at radii smaller that $\rcav$, leading to ground mode frequencies $\wo$ larger than $\wq(\rcav)$. From Figure \ref{fig: eccentricity potentials for different values of Z}, we notice that the ground mode frequencies $\wo$ are approximately equal to the maximum of the eccentricity potential.

To have a broader overview of the behavior of the ground modes we plot in Figure \ref{fig: eccentricity potentials for different values of wp/wq} the eccentricity potentials (for a fixed density profile) for various values of the pressure-to-gravity induced cavity precession ratio $\wp/\wq$. For the sake of visualizing both very low and high values of $\wp/\wq$ in the same graph, in Figure \ref{fig: eccentricity potentials for different values of wp/wq}, we scale $\omecc$ by the sum of pressure- and gravity-induced cavity precession frequencies $\wq+\wp$. For $\wq+\wp\lessapprox 10^{-2}$, the location of the potential well changes significantly with the change in the value of $\wp/\wq$, as seen above. However, for thick disks, $\wp\gg\wq$, pressure support dominates at $\rcav$ effectively anchoring the peak of the eccentricity potential there. Even though the height of the eccentricity potential increases with larger $\wp$, comparing the top panel of Figure \ref{fig: frequency solutions for 3 density models - 2D} to Figure \ref{fig: eccentricity potentials for different values of wp/wq}, we see that the frequency of the ground mode $\wo$ is much smaller than the maximum of the eccentricity potential. This is in line with strong pressure forcing having a stabilizing effect.

\subsubsection{No Solutions for 2D Power-law Disks with $\wp \gg \wq$}\label{subsection: No Solutions for 2D Power-law Disks with high wp_wq ratio}

In Section \ref{sec:3}, we found the eccentricity potential (Eq. (\ref{eq: ecc pot power})). For the power-law density disk's eccentricity potential to be able to trap modes, it must be positive. Therefore, we find the highest value of the $\wp/\wq$ ratio for which the eccentricity potential well exists by finding the value of the $\wp/\wq$ ratio for which the eccentricity potential is equal to zero at the inner radius (see \S \ref{subsec: Disk Density Profiles Allowing Trapped Modes}). 

By Eq. (\ref{eq: nus}), for a 2D locally isothermal ($\xi=0$) disk, and the $\Sigma_1$ density profile ($\ss=-1/2$), the value of the Bessel $\nu$ parameter is $\nu=5/4$. For that $\nu$ value, the eccentricity potential is positive near the inner edge, \ie the inequality (\ref{eq:condition on disk parameters for modes}) is satisfied, only for $\wp<0.76\,\wq$. The condition $\wp<0.76\,\wq$ agrees with the $\wp/\wq$ value, above which no modes are allowed for a 2D locally isothermal disk with the $\Sigma_1$ density profile in Figure \ref{fig: frequency solutions for 3 density models - 2D}.

There is no such limiting $\wp/\wq$ value for the density profiles with an exponential-shaped cavity (Figure \ref{fig: frequency solutions for 3 density models - 2D}). To understand why that is the case, we compare Figure \ref{fig: eccentricity potentials for different values of Z} to Figure \ref{fig:enter-label}. The derivative of the density profile $\Sigma_2$ changes smoothly from approximately zero value at $r=\ab$, to a large positive value at $r\approx\rcav$ (top panel of Figure \ref{fig: eccentricity potentials for different values of Z}). Thus, there is a radius $r\lessapprox \rcav$ where the density profile can be approximated with a power-law function that has a slope with value between $1$ and $3$. Looking at Figure \ref{fig:enter-label}, we see that this slope results in a positive pressure eccentricity potential in that area, thus creating an eccentricity potential well, allowing trapped modes. For large $\wp/\wq$ ratios, the pressure eccentricity potential is the dominant eccentricity potential contribution, and the location of the potential well is determined by the density profile, in the way explained above. We see this in Figure \ref{fig: eccentricity potentials for different values of wp/wq}, where, for $\wp>\wq$, the magnitude of the eccentricity potential changes with changing the value of $\wp/\wq$, but its general location stays the same.

\subsubsection{The Number of Eccentricity Modes}
\label{subsection: Disk Density Impact on Eccentricity - The number of modes}

\begin{figure}[t!]
    \centering
    \includegraphics[scale=1]{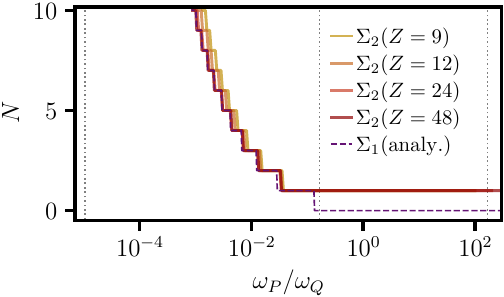}
    \caption{The number of possible modes $N$ in 2D locally isothermal disks. Approximate analytical solutions using power-law density profiles ($\Sigma_1$, dashed purple) are given by Eq. (\ref{eq: number of modes}). Numerical solutions for density profiles with central cavities ($\Sigma_2$, orange) are plotted for different values of the cavity steepness parameter $Z = \{9,12,24,48\}$, as a function of cavity induced frequency ratio $\wp/\wq$. The vertical gray dotted lines mark the three reference values detailed in Table \ref{table:cav_ratio}. }    
    \label{fig: the number of modes N- w_p/w_q for different values of Z}
\end{figure}

We compute the number of possible eccentricity modes for disks with a central exponential cavity to show that the results deviate only slightly from the analytical results for power-law density profiles found in Section \ref{sec:3a}. In Figure \ref{fig: the number of modes N- w_p/w_q for different values of Z}, we plot the number of modes $N$ as a function of the pressure-to-gravity induced cavity precession ratio $\wp/\wq$ in 2D locally isothermal disks ($\xi=0$, $\gamma=1$) for $\Sigma_1$ and $\Sigma_2$, for the same set of parameters as in Figure \ref{fig: frequency solutions for 3 density models - 2D} and Figure \ref{fig: eccentricity potentials for different values of Z}. For each value of $Z$ and $\wp/\wq$, we find the maximum value of the eccentricity potential $\max{(\omecc)}$ required to solve the problem numerically (see \S\ref{S:Methods_numerics}).

For comparison, in Figure \ref{fig: the number of modes N- w_p/w_q for different values of Z}, we plot the approximate, analytical solution for $N$ given by Eq. (\ref{eq: number of modes}) for a disk with power-law density profile with a sharp cavity (Eq. (\ref{eq: sigma_1})). Just like the ground mode frequency solutions $\wo$ (Figure \ref{fig: frequency solutions for 3 density models - 2D}) and the eccentricity potential (Figure \ref{fig: eccentricity potentials for different values of Z}), the number of modes for disks with a central exponential cavity approaches the number of modes for disks with a sharp cavity as $Z$ increases. Similarly, the number of modes obtained for different values of $Z$ are identical for $\wp/\wq\approx10^{-1}$ for all density profiles considered. For low values of $\wp/\wq$, $N$ increases with decreasing $Z$. This is because the effective inner radius decreases (Figure \ref{fig: eccentricity potentials for different values of Z}), and by Eq. (\ref{eq: number of modes}), more modes fit in a wider potential well. Given that $Z$ is the only parameter of the density profile given by Eq. (\ref{eq: sigma_2}) that influences the eccentricity solutions, and its effect on $N$ is not significant (Figure \ref{fig: the number of modes N- w_p/w_q for different values of Z}), we conclude that for the number of modes is fairly insensitive to the choice of density model for $10^{-2} \lessapprox\wp/\wq$ for which modes are possible, and approximately equal to the analytic solution given by Eq. (\ref{eq: number of modes}) for power-law-density profiles.

\subsection{2D Disk Eccentricity for Different Density Profiles}
\begin{figure}[h]
    \centering
    \includegraphics[scale=1]{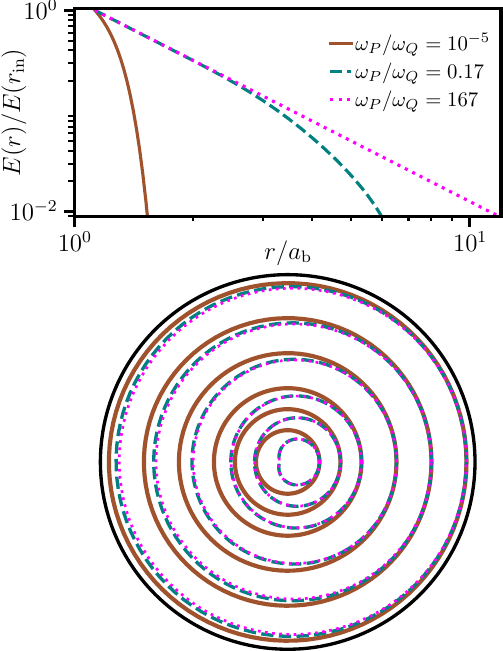}
    \caption{
    Ground mode eccentricity profiles $E(r)$ (top panel) and the disk fluid element orbits (bottom panel)
    for the $\Sigma_2$ density profile and a 2D locally isothermal disk for cavity frequency ratio values detailed in Table \ref{table:cav_ratio}. For each eccentricity profile, we plot orbits (for $r<4\ab$) in the reference frame rotating with the disk precession frequency, assuming an eccentricity profile global scale $E(\rin)=0.2$.}
    \label{fig:ecc_profiles}
\end{figure}

In Figure \ref{fig:ecc_profiles}, we plot eccentricity profiles of ground modes (for $\omega_0$) for three cavity frequency ratios from Table \ref{table:cav_ratio} and $\rin=1.128\ab$. The eccentricity in disks with higher cavity frequency ratios $\wp/\wq$ falls off less steeply with radius, as described in \cite{ML2020}. We note that the eccentricity equation given by Eq. (\ref{eq: eccentricity equation in terms of f_p and f_g})-(\ref{eq: pressure forcing term f_p for the eccentricty equation}) is scale-free, allowing us only to comment on the radial extent of the eccentricity, not it's absolute value. Whether disks with higher cavity induced precession frequency ratios $\wp/\wq$ are more eccentric than disks with lower cavity induced precession frequency ratios will depend on the global scale for the eccentricity profile $E(\rin)$. Therefore, we can not determine the value of $E(\rin)$ within the framework we use in this work.

For eccentricity profiles given in the upper panel of Figure \ref{fig:ecc_profiles}, we plot disk fluid orbits in the bottom panel of Figure \ref{fig:ecc_profiles}. We plot orbits $r(\phi)$ in the reference frame rotating with angular frequency equal to the mode frequency. We scale the eccentricity at the inner edge to $E(\rin)=0.2$ for all three disks, and plot orbits for $r\leq 4 \ab$.

\subsection{Comparing 2D and 3D Disks}
\label{subsec:comparison_cavity_no_cavity}

\begin{figure}[t!]
    \centering
    \includegraphics[scale=0.97]{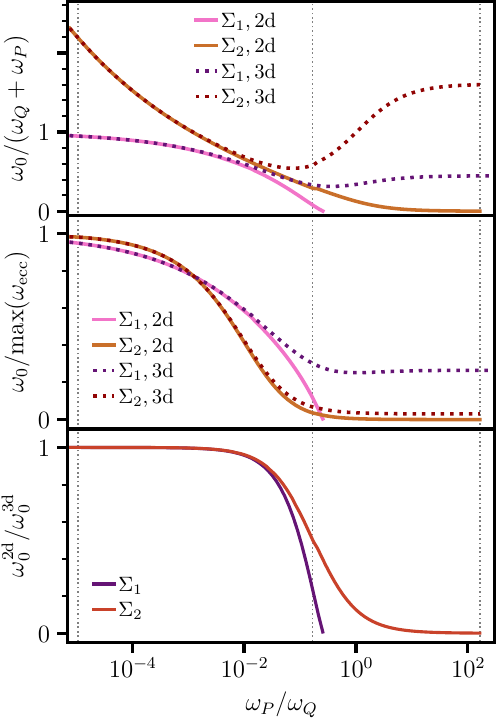}
    \caption{Ground mode frequency solutions $\wo$ for 2D disks (solid lines) and 3D disks (dotted lines) for a power-law-density profile (Eq. (\ref{eq: sigma_1})) (purple lines) and fiducial density profile given by Eq. (\ref{eq: sigma_2}) (orange lines). We plot $\wo$ solutions in units of the sum of quadrupole- and pressure-induced cavity precession frequencies $\wq+\wp$ (upper panel), in units of the maximum value of the eccentricity potential (middle panel), and the ratio od 2D to 3D frequencies (bottom panel). The vertical gray dotted lines mark the three reference $\wp/\wq$ values from Table \ref{table:cav_ratio}. Regardless of the density profile, there is no difference between the 2D and the 3D solutions for razor-thin disks. But for intermediate and very thick disks, the 3D solutions are larger.}
    \label{fig: frequency results 2D-3D comparison}
\end{figure}

We compare the ground mode ($n=0$) frequency $\wo$ for 2D and 3D disks. In Figure \ref{fig: frequency results 2D-3D comparison}, we plot $\wo$ solutions for 2D disks (solid lines) and 3D disks (dotted lines), for power-law-density profiles, Eq. (\ref{eq: sigma_1}) with purple lines, and the fiducial density profile, Eq. (\ref{eq: sigma_2}) with orange lines. Each panel shows $\wo$ normalized by a different quantity, as we explain below.

In the upper panel of Figure \ref{fig: frequency results 2D-3D comparison}, we plot $\wo$ solutions in units of the sum of cavity precession frequencies $(\wq+\wp)$. For both density profiles, the solutions for the 2D and the 3D disks are identical for thin disks for values of the pressure-to-gravity cavity precession ratio $\wp/\wq\lessapprox 10^{-2}$. This is because the dimension of the disk contributes to the eccentricity only through the pressure eccentricity potential and so does not contribute for small values of $\wp/\wq$, for thin disks. Therefore, for thin disks, as expected, there is no difference between 2D versus 3D, and it is just the steepness of the cavity determining $\wo$.

For 3D disks, $\wo/\wq$ solutions start deviating from 2D solutions for $\wp/\wq \gtrsim 10^{-1}$, with the difference growing for larger values of $\wp/\wq$, because the 3D pressure eccentricity potential is larger than the 2D pressure eccentricity potential.

In \S \ref{subsubsection: Disk Density Impact on Eccentricity - The eccentricity potential as an explanation for the solution dependence on cavity steepness}, we showed that the peak value of the eccentricity potential well increases as $Z$ decreases. These facts motivate presenting $\wo$ solutions in a third way, in units of the maximum of the total eccentricity potential within the disk, which we do in the bottom panel of Figure \ref{fig: frequency results 2D-3D comparison}. For $\wp/\wq\ll1$, solutions for both density profiles and both 2D and 3D disks are in a very good agreement. This indicates that for all density profiles with or without a central exponential cavity, the solution for the precession frequency of the lowest order mode $\wo$ is approximately equal to the maximum of the eccentricity potential for razor-thin disks (Figure \ref{fig: eccentricity potentials for different values of Z})

Most numerical hydrodynamical simulations of circumbinary disks simulate systems with the $\wp/\wq$ value such that we expect the 3D disk precession frequency to be almost twice as large as the 2D disk precession frequency (center dashed gray vertical line in Figure \ref{fig: frequency results 2D-3D comparison}, Table \ref{table:cav_ratio}). This ratio is in agreement with comparison of 2D and 3D cavity precession rates found in isothermal hydrodynamical simulations for binary+disk systems with the same scale height, mass ratio, and binary eccentricity \citep[Figure 11 of][]{KITP_CC:2024}.

\section{Discussion and Conclusion}
\label{sec: Discussion and conclusion}
We carried out a semi-analytical, perturbative approach to study eccentric CBDs. This allowed us to characterize solutions for 
eccentric disk modes, namely through the distribution of the eccentricity of disk fluid elements $E(r)$, the precession frequency $\omega$, the mode existence and multiplicity, and the solution dependence on binary and disk parameters, as well as modeling choices.

To keep track of how the disk dimension and its equation of state influences solutions, we first generalized the eccentricity perturbation equation (Section \ref{sec:methods}). To do so we re-derived the equations describing the evolution for eccentric disks \citep{Ogilvie_2001,TO2016}, and kept track of the temperature perturbations by making them proportional to the adiabatic temperature perturbation, and a labeling parameter $\xi$. Setting $\xi=0$ is equivalent to requiring no perturbations with respect to the initial radial temperature profile, making the disk locally isothermal. Setting $\xi=1$ allows for the temperature perturbation necessary for disk fluid elements to be on adiabatic orbits.

To examine how the choice of the density profile influences the disk eccentricity solutions, we considered several different density profiles. We started with power-law-density profiles in Section \ref{sec:3}, for which we conducted an analytical analysis of eccentricity solutions in non-precessing (\S \ref{sec:3a}) and precessing (\S \ref{sec:3b}) disks. We end our analysis with Section \ref{sec: Disk Density Impact on Eccentricity}, in which we found numerical solutions for disks with density profiles that account for viscous forces, boundary effects, and a smooth central cavity.  

Finally, the solutions to the eccentricity equation depend on the value of the disk aspect ratio $h$, and the strength of the binary quadrupole gravitational potential set by the binary mass ratio $q$ (Eq. (\ref{eq: definition of the Q parameter})). As \cite{ML2020} showed numerically, and as discussed in \S \ref{sec:3}, the quantity through which these values appear in all solutions is the pressure-to-gravity induced frequency ratio at the cavity radius (Eq. (\ref{eq:omega_PQ})). Specifically, for a chosen density profile, the solution for $\wo/(\wp+\wq)$ is uniquely determined only by the ratio of cavity induced precession frequencies $\wp/\wq$. However, the value of the precession frequency $\wo$ does change with changing the values of $\wq+\wp$, even if the ratio $\wp/\wq$ stays the same. This is the case because it is proportional to $(\wp+\wq)$.
Throughout the paper, we highlighted results for three reference values of the pressure-to-gravity frequency ratio (Table \ref{table:cav_ratio}).  Here we discuss the main differences in results between the three regimes that these reference ratio values represent.

\paragraph{Very thick disks}\label{Very thick disks}
The reference value $\wp/\wq \approx 167$ is representative of a very thick disk around a small mass ratio binary. In this discussion, when referring to very thick disks, we consider $ \wp/\wq\gtrapprox 10$. For an equal mass binary and $\rcav=2.5\ab$, this corresponds to the disk aspect ratio $ h\gtrapprox0.8$. For a disk aspect ratio $h=0.15$ and $\rcav=2.5\ab$, this is equivalent to the binary mass ratio of $q\lessapprox0.01$. 
The former case is not of interest, as we do not expect the accretion disks around black holes \citep[\eg,][]{Hubeny_2001} or circumstellar disks \citep[\eg][]{Wood_1999} to be that thick, and circumbinary disks with that aspect ratio are likely not truncated even by equal mass ratio binaries. 
The latter scenario includes a binary mass ratio too small to clear a low-density annulus or cavity \citep{Duffel_2015},
contradicting the assumption of a cavity we use throughout this paper. Consequently, we conclude that the very thick disk range as modeled in this work does not represent feasible scenarios for circumbinary disks.

\paragraph{Intermediate disks}
The reference value $\wp/\wq \approx 0.17$ is representative of most numerical hydrodynamical simulations of accreting circumbinary disks, which simulate disks with the disk aspect ratio in range $0.05\leq h \leq 0.1$. In this discussion, when referring to intermediate disks, we consider $10^{-2} \lessapprox \wp/\wq\lessapprox 1$. For an equal mass binary and $\rcav=2.5\ab$, this corresponds to the disk aspect ratio $0.025 \lessapprox h\lessapprox0.25$. 
Notably, for intermediate disks, a maximum of one solution is possible across all disk types we considered in this work (Figure \ref{fig: the number of modes N- w_p/w_q for different values of Z}), providing a possible explanation as to why higher modes were not yet observed in simulations. For intermediate disks and an exponential cavity, the precession frequency solutions are insensitive to the details of the cavity shape (bottom panel of Figure \ref{fig: frequency solutions for 3 density models - 2D}). This could serve as a possible explanation for why this is a robust result in the simulation literature, most of which use disks in this regime \citep{ML2020,Sudarshan_2022, KITP_CC:2024}. The precession frequency for 3D disks is larger than for 2D disks. The difference between the 3D and 2D solutions increases with increasing disk aspect ratio or decreasing binary mass ratio. For the ($h=0.1$, $q=1$) system, the precession frequency of a 3D disk is almost twice the magnitude of the 2D disk frequency. This is in good agreement with numerical solutions for 2D and 3D disks simulated by multiple codes for the $h=0.1$, $q=1$ problem in \citep[][]{KITP_CC:2024}.

\paragraph{Razor thin disks}
The reference value $\wp/\wq \approx 10^{-5}$ is representative of disks around black holes, which are likely razor-thin. In this discussion, when referring to razor thin disks, we consider $\wp/\wq\lessapprox10^{-2}$. For an equal mass binary and $\rcav=2.5\ab$, this corresponds to the disk aspect ratio $h\lessapprox0.025$. Unlike for thicker disks, the precession frequency of a razor-thin eccentric disk is very sensitive to the shape of the central cavity, namely its location and steepness. In thin disks, the solution for the frequency is mostly influenced by the gravitational potential of the binary, which is largest at the inner edge of the disk.
Hence, for razor-thin disks, the precession frequency is larger if the cavity is smaller and shallower, because the gravitational eccentricity potential is stronger closer to the binary. Therefore, disks with smaller or shallower cavities have a larger gravitational potential at their inner edge and exhibit higher precession frequencies. In razor thin disks, the ground mode precession frequency is approximately equal to the maximum of the eccentricity potential, whose value is very sensitive to the cavity shape. In practice, to quickly estimate the value of the precession frequency of thin disks, we need only a good estimate of the cavity shape, while the density profile at larger radii is of no significance. This suggests that predictions for the cavity truncation shape, with disk and binary parameters, could allow one to predict disk eccentricity and precession properties. 
The dimension, 2D versus 3D, of the disk is present in the problem only through pressure, which is negligible in thin disks, making the frequency solution for thin disks independent of the disk dimension. Thin disks can support many modes, possibly making their evolution much more complicated. Higher order modes, however, have not been reported in hydrodynamical simulations of thin disks \citep{Tiede+_2024}, though they have not been explicitly searched for. The number of modes is proportional to the binary mass ratio and the inverse of the disk aspect ratio, and is fairly insensitive to the choice of the density profile. This makes the analytic expression found in Section \ref{sec:modes} a good approximation for razor-thin disks, at least in the examined $10^{-4} \lessapprox \wp/\wq \lessapprox 10^{-1}$ range. For razor thin disks, the ground mode is highly localized near the cavity, so the disk is eccentric only there. Higher-order modes spread further out, making it possible for significant values of the eccentricity to be found up to the radius of tens or hundreds of binary separations (Figure \ref{fig: eccentricity profiles for non-precessing disks - 3 panels}). Higher-order modes have been presented as possible solutions to the eccentricity equation \citep{TO2016,ML2020}, but it is unclear what would excite a higher-order mode, or what would its amplitude be compared to the ground mode.

In this work, we discuss circular orbit binary systems ($\eb=0, q\neq0$). Additionally, we focus on thin disks, for which we expect the binary with a mass ratio larger than $q=0.05$ to clear out a central cavity in the disk (see discussion in Section (\ref{subsection: Disk Density Impact on Eccentricity - Three Useful Density Models})). Furthermore, in Section \ref{sec:3} and Section \ref{sec: Disk Density Impact on Eccentricity}, we have restricted our analysis to disks with a constant disk aspect ratio, but the perturbative approach \citep{TO2016} we present in Section \ref{sec:methods} and Section \ref{sec:3} is applicable to other choices. It is also applicable to a disk around a single black hole. For example, \cite{Ogilvie:2008} applied the perturbative approach to a non-constant disk aspect ratio ($h\propto r^{1/2}$) and for $Q=0$. \cite{Saini_Gulati_Sridhar_2009} also study pressure modes for disks around single black holes, including non-localized solutions. Additionally, one could include the self-gravity of the disk, as considered by \cite{Papaloizou_2002,Lee_Dempsey_Lithwick_2019}.

In summary, we applied a systematic approach to study circumbinary disk eccentricity solutions, using perturbative methods. 
We presented analytical solutions for power-law density profiles that we used to find estimates for more complex density models.
We have found that, for razor-thin disks, such as the ones we expect to form around MBHBs \citep{Shakura_Sunyaev_1973,Krolik_1999}, the disk precession frequency is approximately equal to the gravitational eccentricity potential evaluated at the cavity edge (sensitive to the cavity shape, determined by the cavity size $\rcav$ and the steepness of the central cavity $Z$), which allows for a quick estimate of its value without solving the eccentricity boundary value problem. Moreover, we provided insight into the number of possible modes, which seems to be fairly insensitive to the choice of the density profiles in this work, but depends significantly on $\wp/\wq=h^2\rcav^2/Q\ab^2$.
In the future, we will apply the analysis and results presented in Section \ref{sec:3} to scenarios with forced eccentricity and facilitate their numerical exploration.

\acknowledgments
M.G. and D.J.D. acknowledge support from the Danish Independent Research Fund through the Sapere Aude Starting Grant No. 121587. 
D.J.D. received funding from the European Union's Horizon 2020 research and innovation programme under Marie Sklodowska-Curie grant agreement No. 101029157.
M.E.P. gratefully acknowledges support from the Independent Research Fund Denmark via grant ID 10.46540/3103-00205B and the Institute for Advanced Study, where part of this work was carried out. The Tycho supercomputer hosted at the SCIENCE HPC center at the University of Copenhagen was used to support this work.

\bibliographystyle{apj} 
\bibliography{refs}
\appendix{}
\label{app1}

\section{The connection between different forms of the  eccentricity equation }
\label{app B - general eccentricty potential and bessel calculations}

We write the eccentricity equation in the following form:
\begin{equation}\label{z1}
    E''+E' f_{E'}+Ef_{E}=0.
\end{equation}
We use the eccentricity equation in the form given by Eq. (\ref{z1}) for our chosen numerical IVP solver (see \S \ref{S:Methods_numerics}), to calculate Schrödinger form parameters (see \S \ref{subsection: Schrödinger form of the eccentricity equation and the eccentricity potential}), and to find the connection between Bessel parameters (Section \ref{sec:3}) and the eccentricity potential in \S \ref{appB - subsection: Bessel parameters}. 

 We combine Eqs. (\ref{eq: eccentricity equation in terms of f_p and f_g})-(\ref{eq: pressure forcing term f_p for the eccentricty equation}) to find the expression that multiplies the first radial derivative of the eccentricity profile $f_{E'}$:
\begin{equation}\label{eq: f_E'}
    f_{E'}=\frac{\left[ r^3 \Sigma^{1-\xi}p^{\xi}\right]'}{r^3 \Sigma^{1-\xi}p^{\xi}},
\end{equation}
as a radial function determined by the density (pressure) radial profile, which retains its form for both 2D and 3D disks.

We combine Eqs. (\ref{eq: eccentricity equation in terms of f_p and f_g})-(\ref{eq: pressure forcing term f_p for the eccentricty equation}) to find the expression that multiplies the eccentricity profile $f_{E}$:
\begin{subequations}\label{eq: f_E}
    \begin{align} \label{eq: 2D f_E}
    \begin{split}
        f_E^{\mathrm{(2D)}}\equiv & \frac{1}{\gamma^{\xi}} \left[ \frac{p'}{rp}-(1-{\xi})\frac{\left[r^3\Sigma(p/\Sigma)'\right]'}{r^3p}+\frac{\Omega^2 \Sigma Q}{p(r/\ab)^2}-\frac{2\Sigma \Omega \omega}{p}\right],
        \end{split}\\
         \label{eq: 3D f_E}
        \begin{split}
        f_E^{\mathrm{(3D)}}\equiv &\left(\frac{\gamma}{2\gamma-1} \right)^\xi\left[ \frac{p'}{rp}\left(\frac{4\gamma-3}{\gamma} \right)^\xi+\frac{3}{r^2}\left(\frac{2\gamma+1}{\gamma} \right)^\xi-\frac{(1-\xi)}{p}\left[ p\frac{(p/\Sigma)}{p/\Sigma}\right]'+\frac{\Omega^2 \Sigma Q}{p(r/\ab)^2}-\frac{2\Sigma \Omega \omega}{p}\right],
        \end{split}
    \end{align}
\end{subequations}
which is determined by the binary quadrupole gravitational strength, disk precession frequency, the disk density (pressure) radial profile, and the disk dimension.

\subsection{Bessel parameters}
\label{appB - subsection: Bessel parameters}
We assume that the density and the pressure are some powers of radius:
\begin{align}\label{eq:d0}
    \Sigma & =\Sigma_0 (r/\ab)^{\ss},\\
    p&=p_0(r/\ab)^{\mathrm{s_p}},
\end{align}
in which case we can write the disc aspect ratio $h^2=\Omega^2r^2\Sigma/p$ as:
\begin{equation}\label{eq:h}
    h=h_0(r/\ab)^{\mathrm{s_H}},
\end{equation}
where $h_0$ is a constant, and $\mathrm{s_p}=\ss-1-2\mathrm{s_H}$.
For power-law density and pressure profiles (Eq. (\ref{eq:d0})), we can write the eccentricity equation (Eq. (\ref{z1})) for a non precessing ($\omega=0$) disk in the form of the Bessel equation \citep{Watson_1922}:
\begin{equation}\label{z2}
    E''+E'\frac{-2\delta_1+1}{r}+E\left[\rs^{-2\delta_2}\delta_2^2r^{2\delta_2-2}+\frac{\delta_1^2-\nu^2\delta_2^2}{r^2} \right]=0, 
\end{equation}
whose solutions are:
\begin{equation}
    E=\beta_1 r^{\delta_1}J_\nu\left((r/\rs)^{\delta_2}\right) +\beta_2 r^{\delta_1}J_{-\nu}\left((r/\rs)^{\delta_2} \right),
\end{equation}
where $\beta_1$ and $\beta_2$ are constants, whose value we find by applying boundary conditions (see Section \ref{sec:methods}).

Substituting Eq. (\ref{eq:d0})-(\ref{eq:h}) into Eq. (\ref{eq: f_E'}), and comparing it to Eq. (\ref{z2}), we find:
\begin{equation}\label{eq:delta_1_general}
    \delta_1=-1+\frac{\xi-\ss}{2}+\xi\mathrm{s_H}.
\end{equation}

Substituting Eq. (\ref{eq:d0})-(\ref{eq:h}) into Eq. (\ref{eq: f_E}), and comparing it to Eq. (\ref{z2}) , we find, for a non precessing ($\omega=0$) disk, the constant $\delta_2$:
\begin{equation}\label{eq: delta_2_general}
    \delta_2=\mathrm{s_H}-1,
\end{equation}
the radial scale $\rs$:
\begin{equation}\label{eq: rs general}
    \rs=\ab \left( \frac{Q}{\gamma_\mathrm{eff}^2h_0^2\delta_2^2}\right)^{-1/2\delta_2},
\end{equation}
and the $\nu$ parameter for 2D and 3D disks:
\begin{subequations}\label{eq: nus}
    \begin{align} \label{eq: 2D nu}
    \begin{split}
        \left(\nu^{2}\right)^\mathrm{2D}\equiv & \frac{1}{\delta_2^2}\left[\delta_1^2- \frac{1}{\gamma^\xi} \frac{s_\mathrm{p}-(1-\xi)(s_\mathrm{p}-\ss)(2+s_\mathrm{p})}{1} \right],
        \end{split}\\
         \label{eq: 3D nu}
        \begin{split}
        \left(\nu^{2}\right)^\mathrm{3D}\equiv &\frac{1}{\delta_2^2}\left[\delta_1^2- \left(\frac{\gamma}{2\gamma-1} \right)^\xi\left( \left(\frac{4\gamma-3}{\gamma} \right)^\xi s_\mathrm{p}+\left(\frac{2\gamma+1}{\gamma} \right)^\xi 3-(1-\xi)\left(\left(s_\mathrm{p}-\ss\right)\left(s_\mathrm{p}-1\right)\right) \right)\right] .
        \end{split}
    \end{align}
\end{subequations}

\paragraph{Disks with constant disk aspect ratio}
 If the disc aspect ratio is a constant ($\mathrm{s_H}=0$), then, by Eq. (\ref{eq: delta_2_general}) $\delta_2=-1$. Expressions for $\delta_1$ (Eq. (\ref{eq:delta_1_general})), $\rs$ (Eq. (\ref{eq: rs general})), and $\nu$ (Eq. (\ref{eq: nus})) reduce to those given in Section \ref{sec:3a}. 

\subsubsection{The eccentricity potential}
To find the connection between the Bessel parameters and the eccentricity potential, we write the wave number function $k^2(r)$ for the eccentricity equation given by Eq. (\ref{z2}). First, we compare Eq. (\ref{z1}) to Eq. (\ref{z2}) to identify $f_{E}=\rs^{-2\delta_2}\delta_2^2r^{2\delta_2-2}+(\delta_1^2-\nu^2\delta_2^2)/r^2 $ and $f_{E'}=(1-2\delta_1)/r$. We then use the relation between the wave number and the functions $f_{E}$ and $f_{E'}$ \citep{Lanczos_1997}:
\begin{equation}\label{eq: k squared def}
    k^2=f_E-\frac{f'_{E'}}{2}-\frac{f^2_{E'}}{4},
\end{equation}
to write the wave number function in terms of Bessel parameters:
\begin{equation}\label{eq: k-nu relationship}
\begin{split}
    k^2&=\rs^2r^{-4}+\frac{\delta^2-\nu^2}{r^2}+\frac{(1-2\delta)}{2r^2}-\frac{(1-2\delta)^2}{4r^2}=\rs^2r^{-4}-\left(\nu^2-\frac{1}{4}\right)r^{-2}.
    \end{split}
\end{equation}

Comparing Eq. (\ref{eq: wave number k^2 for Schrödinger equation}) to Eq. (\ref{eq: k-nu relationship}), we find the eccentricity potential $\omega_\mathrm{ecc}$ expressed in Bessel parameters:
\begin{equation}
        \omega_\mathrm{ecc}=\frac{\Omega h_e^2}{2}\left[ \frac{\rs^2}{r^2}-\left( \nu^2-\frac{1}{4}\right)\right].
\end{equation}

\subsection{Schrödinger form}
\label{appB - subsection: Schrödinger form}

We can write Eq. (\ref{z1}) as a Schrödinger equation $y''+k^2y=0$ for the scaled eccentricity $y$. The relationship between the eccentricity $E(r)$ and the scaled eccentricity $y(r)$ is given by \citep{Lanczos_1997}:
\begin{equation}\label{eq: scaled ecc def}
    y\equiv E \exp{\left[\frac{1}{2}\int f_{E'}dr\right]}.
\end{equation}
We substitute $f_{E'}$ from Eq. (\ref{eq: f_E'}) into Eq. (\ref{eq: scaled ecc def}), to find the scaled eccentricity:
\begin{equation}\label{eq: scaled ecc}
    y=Er^{3/2}p^{\xi/2}\Sigma^{(1-\xi)/2}.
\end{equation}

To find $k(r)$, we substitute $f_{E'}$ from Eq. (\ref{eq: f_E'}) and $f_{E}$ from Eq. (\ref{eq: f_E}) into Eq. (\ref{eq: k squared def}), and present results in \S \ref{subsection: Schrödinger form of the eccentricity equation and the eccentricity potential}.

\paragraph{Trapped modes and the choice of inner and outer radius for density profiles with a smooth cavity}
In Section \ref{sec:3}, the power-law-density profiles described a non-zero density from $\rcav<r<\infty$, so it was natural to choose $\rin=\rcav$ for the inner disk radius used to solve equations presented in Section \ref{sec:methods}. But for density profiles with a smooth cavity, such as $\Sigma_2$ and $\Sigma_3$, the choice of $\rin$ isn't as clear. For these density profiles, however, the modes are trapped within the potential well (see bottom panel of Fig \ref{fig: eccentricity potentials for different values of Z}). We once again use quantum mechanics analogy to conclude that the exact values of $\rin$ and $\rout$ should not affect trapped mode solutions as long as they are outside of, and far enough from, the potential well \citep{ML2020}. Since the location of the potential well changes rapidly when changing the values of $\wp/\wq$ and $Z$ for razor-thin disks and is different for different cavity sizes $\rcav$, we cannot set a unique value $\rin$ that would be suitable for all parameters used in this work. Furthermore, decreasing the value of $\rin$, and increasing the value of $\rout$ drastically increases the computation time. To tackle both issues, we do the following. For every combination of parameters ($Z,\rcav,L,\wp/\wq$), we find the eccentricity potential and locate its first null point $r_1$, which serves as the lowest possible left turning point of a trapped mode. We set the inner radius $\rin$ to be some radius smaller that $\rin$, and the outer integration radius
$\rout$ to be proportional to the value of $\wp/\wq$.

\end{document}